\documentstyle[12pt]{article}
\input{psfig.sty}
\begin{document}

\begin{titlepage}


\begin{center}
\baselineskip 24pt
{\Large {\bf CKM Matrix and Fermion Masses in the Dualized Standard Model}}\\
\vspace{.5cm}
\baselineskip 14pt 
{\large Jos\'e BORDES}\\
bordes\,@\,evalvx.ific.uv.es\\
{\it Dept. Fisica Teorica, Univ. de Valencia,\\
  c. Dr. Moliner 50, E-46100 Burjassot (Valencia), Spain}\\
\vspace{.2cm}
{\large CHAN Hong-Mo, Jacqueline FARIDANI}\\
chanhm\,@\,v2.rl.ac.uk \quad faridani\,@\,hephp1.rl.ac.uk\\
{\it Rutherford Appleton Laboratory,\\
  Chilton, Didcot, Oxon, OX11 0QX, United Kingdom}\\
\vspace{.2cm}
{\large Jakov PFAUDLER}\\
jakov\,@\,thphys.ox.ac.uk\\
{\it Dept. of Physics, Theoretical Physics, University of Oxford,\\
  1 Keble Road, Oxford, OX1 3NP, United Kingdom}\\
\vspace{.2cm}
{\large TSOU Sheung Tsun}\\
tsou\,@\,maths.ox.ac.uk\\
{\it Mathematical Institute, University of Oxford,\\
  24-29 St. Giles', Oxford, OX1 3LB, United Kingdom}\\
\end{center}

\begin{abstract}
A Dualized Standard Model recently proposed affords a natural explanation
for the existence of Higgs fields and of exactly 3 generations of fermions,
while giving at the same time the observed fermion mass hierarchy together
with a tree-level CKM matrix equal to the identity matrix.  It further suggests 
a method for generating from loop corrections the lower generation masses 
and nondiagonal CKM matrix elements.  In this paper, the proposed calculation 
is carried out to 1-loop.  It is found first that with the method suggested
one can account readily for the masses of the second generation fermions
as a `leakage' from the highest generation.  Then, with the Yukawa couplings
fixed by fitting the masses of the 2 higher generations, one is left with
only 2 free parameters to evaluate the CKM matrix and the masses in the 
lowest generation.  One obtains a very good fit to the CKM matrix and 
sensible values for the masses of $d$ and $e$, though, for a valid reason, 
not of $u$.  In addition, the fitted values of the Yukawa couplings and 
vacuum expectation values of the dual colour Higgs fields show remarkable 
features perhaps indicative of a deeper significance.

\end{abstract}

\end{titlepage}

\clearpage

\setcounter{section}{0}
\setcounter{equation}{0}
\def\theequation{\arabic{section}.\arabic{equation}}

\section{Introduction}

Up to the present, the Standard Model has worked exceptionally well, there
being no experimental fact we know which is demonstrably contradictory to
its predictions.  Nevertheless, the Standard Model contains in itself a 
number of unsatisfactory features, which are widely recognized as such.  
For example, at the more fundamental level, Higgs fields are introduced to
break the electroweak symmetry and fermions are assumed to exist in three
generations or families to fit observation without theoretical reasons 
being given for why it should be so.  Compared with the intrinsic gauge 
structure and the existence of the gauge bosons in the theory, both of 
which have deep geometric significance, the assumptions about Higgs fields 
and fermion generations appear {\it ad hoc}.  At the more practical level, this
situation is reflected in the large number of independent parameters which
have to be determined by experiment.  Besides, these parameters exhibit 
some quite startling patterns which are still unexplained.  In particular,
there is first the so-called fermion hierarchy puzzle, namely that fermions 
of the same type but different generations have widely different masses.  
Take, for example, the three $U$-type quarks; the experimental values quoted 
in the latest data booklet \cite{databook} for the masses of $t, c$, and $u$ 
respectively are 176 $\pm$ 5 GeV, 1.0 - 1.6 GeV, and 2 - 8 MeV, dropping 
by more than two orders of magnitude from generation to generation.  Then 
secondly, there is the mixing problem, say, between the $U$-type and $D$-type 
quarks through the CKM matrix \cite{CKM,Jarlskog}, which though tantalisingly 
close to the identity matrix is yet not the identity, with its off-diagonal 
elements varying in magnitude from about 20 percent to about 3 permille 
\cite{databook}.  These empirical facts, of course, are all of the greatest 
phenomenological significance and cry out for a theoretical explanation 
but are not given one in the Standard Model as usually formulated.  

In the literature, answers to these questions are often sought for from 
beyond the Standard Model, but with, to our minds, no obvious great 
success.  The difficulty is that, there being more freedom working 
outside the Standard Model framework, one often ends up by putting in 
more than one gets out.  Recently, however, a suggestion was made for a 
solution of the above problems from within the framework of the Standard 
Model itself which, if at all possible, would at least have the advantage 
of economy and restraint.  In this suggestion \cite{Chantsou}, one first 
made use of a newly discovered generalized electric-magnetic duality for 
Yang-Mills fields \cite{Chanftsou} together with a well-known result of 
't~Hooft on confinement \cite{thooft} to give a natural place to both 
Higgs fields and fermion generations, with Higgs fields appearing as 
frame vectors in internal symmetry space and fermion generations appearing 
as dual colour.  As an immediate consequence, one then deduced that 
fermions will occur in exactly three generations and that the generation 
symmetry will be broken, as experimentally observed.  Moreover, with one 
more simple assumption about the dual hypercharges of the dual Higgs 
fields, it was shown that there will be a fermion mass hierarchy and 
that the CKM matrix would be the identity at tree level, but that loop 
corrections would lift the above tree-level degeneracy to give small but
nonzero values both to the lower generation fermion masses and to the 
off-diagonal CKM matrix elements.  

The purpose of the present paper is to push further in this direction to 
make a first attempt at actually evaluating quark masses and CKM matrix
elements for comparison with experiment.  The calculation is here carried out
to 1-loop level.  Out of the many 1-loop diagrams we have examined, it
turns out first that some, which affect only the normalization of the
fermion mass matrix but not its orientation, are large due to the large
dual gauge coupling and cannot be evaluated perturbatively.  Since it is
only the orientation in flavour space which is of the most interest to us
as far as the CKM matrix is concerned, it is profitable at present to 
abandon calculating the normalization of the mass matrix and concentrate 
solely on its orientation.  This has the benefit of allowing us to ignore 
those diagrams affecting only the normalization, reducing thus the number 
of free parameters in the problem.  Secondly, it happens that of the 
remaining diagrams affecting the orientation of the fermion mass matrix, 
most are negligible if we put in the estimate for the dual gauge
boson mass obtained from the absence of flavour-changing neutral decays.
As pointed out already in \cite{Chantsou}, the exchange of the dual gauge
bosons would give rise to FCNC effects, and experimental constraints put
a lower bound on the lowest dual gauge boson mass of several 100 TeV.

What remains then is basically just the Higgs loop diagram which matters
for our present investigation.  This depends on a Yukawa coupling strength
$\rho$, one for each fermion type, a mass scale $m_T$ which may be identified
as the highest generation mass, again one for each fermion type, and lastly
the 3 vacuum expectation values $(x, y, z)$ of the dual colour Higgs fields
which are common to all fermion types.  We ascertain first that the masses 
of the second generation fermions can indeed be obtained as a `leakage'
from the highest generation, as suggested in \cite{Chantsou}, with a
Yukawa coupling strength $\rho$ of order unity for each fermion type.
We then fixed their values by fitting these $\rho$'s to the empirical values
of the masses of the second generation.  Next, of the remaining parameters 
$(x, y, z)$, it was shown that the calulation is independent of their 
normalization to a high accuracy.   With then only 2 free parameters, we had
to calculate the CKM matrix and the fermion masses of the lowest generation.
A very good fit to the absolute values of all CKM elements was obtained together
with some ratios and products of these elements measured independently.
In addition, in spite of the lack of knowledge on the scale-dependence 
of the normalization of the mass matrix, sensible estimates were obtained 
for all the lowest generation fermions except for the $u$-quark.  

The result of the fit reveals also 2 intriguing features, namely (a) a close
proximity of the normalized vector $(x, y, z)$ to one of its fixed points 
$(1,0,0)$ to an accuracy of about one part in ten thousand, and (b) the near 
equality to a few percent accuracy of the fitted values of the $\rho$'s for
all 3 fitted fermion types (i.e. U, D and the charge leptons L).  We think 
these may be indicative of a hidden symmetry which we have not yet understood.

\setcounter{equation}{0}

\section{The Framework}

We begin with a resum\'e of the dual framework on which the calculations are
based, the details of which can be found in \cite{Chantsou}.  Generalized
electric-magnetic duality as obtained in \cite{Chanftsou} implies that
dual to colour in the Standard Model, there is also an $\widetilde{SU}(3)$
symmetry for dual colour.  The charges of this dual symmetry are colour 
monopoles and its monopoles are colour charges.  Using then the well-known 
result of 't~Hooft \cite{thooft}, one concludes from the fact that colour 
is confined that dual colour $\widetilde{SU}(3)$ is spontaneously broken 
and Higgsed.  The proposal was that this broken dual colour symmetry be
identified with what is sometimes referred to in the literature as the
``horizontal symmetry'' relating the generations \cite{horizontal,horizontal1}.

Now it so happens that in the dual framework of \cite{Chantsou} there are 
scalar fields occurring which have the right properties to play the role
of Higgs fields, these being the frame vectors in the $\widetilde{SU}(3)$
internal space.  They constitute altogether 3 dual colour triplets, which
we denote by $\phi^{(a)}_{\tilde{a}}$, with $\tilde{a} = 1, 2, 3$ 
representing the dual colour which labels the 3 components of a triplet and 
$(a) = 1, 2, 3$ being just a label distinguishing the 3 triplets.
We want the vacuum expectation values of $\phi^{(a)}_{\tilde{a}}$ to form 
an orthogonal triad, as is appropriate for the 3 vectors which make up 
an $\widetilde{SU}(3)$ frame.  We need therefore a Higgs potential which 
gives these vacuum expectation values as minima.  The following was 
suggested \cite{Chantsou}:
\begin{equation}
V[\phi] = -\mu \sum_{(a)} |\phi^{(a)}|^2 + \lambda \left\{ \sum_{(a)}
   |\phi^{(a)}|^2 \right\}^2 +\kappa \sum_{(a) \neq (b)} |\bar{\phi}^{(a)}
   .\phi^{(b)}|^2,
\label{Higgspot}
\end{equation}
with $\mu, \lambda$ and $\kappa$ all positive.  The minimum of $V$ occurs when 
the $\phi^{(a)}$ are mutually orthogonal and $\sum_{(a)} |\phi^{(a)}|^2 
= \mu/2\lambda$, independently of the individual lengths of the different 
$\phi^{(a)}$'s.  Thus, a vacuum can be chosen as:
\begin{equation}
\phi_V^{(1)} = \zeta \left( \begin{array}{c} x \\ 0 \\ 0 \end{array} \right),
\phi_V^{(2)} = \zeta \left( \begin{array}{c} 0 \\ y \\ 0 \end{array} \right),
\phi_V^{(3)} = \zeta \left( \begin{array}{c} 0 \\ 0 \\ z \end{array} \right),
\label{phivac}
\end{equation}
with
\begin{equation}
\zeta = \sqrt{\mu/2\lambda}, 
\label{zeta}
\end{equation}
and
\begin{equation}
x^2 + y^2 + z^2 = 1,
\label{xyznorm}
\end{equation}
which will in general break both the $\widetilde{SU}(3)$ symmetry and 
the permutation symmetry between the different $\phi^{(a)}$'s.  In our
calculation here we shall use explicitly this potential although we
recognize that it has no claim for uniqueness.  We shall show, however, 
that the result will not depend much on its detailed properties.

As in \cite{Chantsou}, left-handed fermions are assigned to dual colour
triplets and right-handed fermions to dual colour singlets.  We have thus 
the Yukawa coupling term:
\begin{equation}
\sum_{[b]} Y_{[b]} \sum_{(a)} (\bar{\psi}_L)^{\tilde{a}} \phi^{(a)}_{\tilde{a}}
   (\psi_R)^{[b]} + h.c.,
\label{Yukawa}
\end{equation}
where we have suppressed both colour and weak isospin indices which are
irrelevant for our discussion here.  Inserting then the vacuum expectation 
values given in (\ref{phivac}) for the Higgs fields, we have at tree-level
the following factorized fermion mass matrix:
\begin{equation}
m = \zeta \left( \begin{array}{c} x \\ y \\ z \end{array} \right) (a, b, c),
\label{fermassmat}
\end{equation}
where we have abbreviated the Yukawa couplings
$Y_{[1]}=a,Y_{[2]}=b,Y_{[3]}=c$.
The matrix being of rank 1, it follows that $m m^{\dagger} $ has only one 
non-zero eigenvalue \cite{Fritzsch}, namely $\rho^2 \zeta^2$, with
\begin{equation}
\rho = \sqrt{|a|^2 + |b|^2 + |c|^2}
\label{rho}
\end{equation}
implying thus immediately a mass hierarchy with one fermion state much 
higher in mass than the other two.  Furthermore, the first factor in $m$ 
being given in terms just of the Higgs vacuum expectation values $x, y, z$, 
is independent of the fermion type, i.e. of whether it is $U$-type or 
$D$-type quarks or leptons that we are dealing with, although the second 
factor, given in terms of the Yukawa couplings $a,b$ and $c$, 
does depend on the fermion type.  As a consequence, one 
obtains that the CKM matrix, which depends only on the relative orientation 
of the first (left-handed) factors of respectively the $U$-type and 
$D$-type quarks, is at tree-level just the identity matrix.  This was 
already discussed in detail in \cite{Chantsou}.

What we need to do now is to go beyond the tree level and look at loop 
corrections.  As pointed out in \cite{Chantsou}, because of the special
manner in which the fermions here are coupled to the dual gauge and 
Higgs bosons, loop corrections do not destroy the factorizability property
of the tree-level mass matrix.\footnote{As a result, the mass matrix has
two zero eigenvalues so that any $\theta$-vacuum can be rotated away and
the strong CP problem is avoided. \cite{Pfaudler}}  Nevertheless, they
will modify the first (left-handed) factor in (\ref{fermassmat}) and 
hence give rise to a nontrivial CKM matrix and nonzero masses to the 
lower generations fermions as explained in \cite{Chantsou}.  This is what 
we wish now to examine in detail.

\setcounter{equation}{0}

\section{One-Loop Diagrams}

\begin{figure}[htb]
\vspace{1cm}
\centerline{\psfig{figure=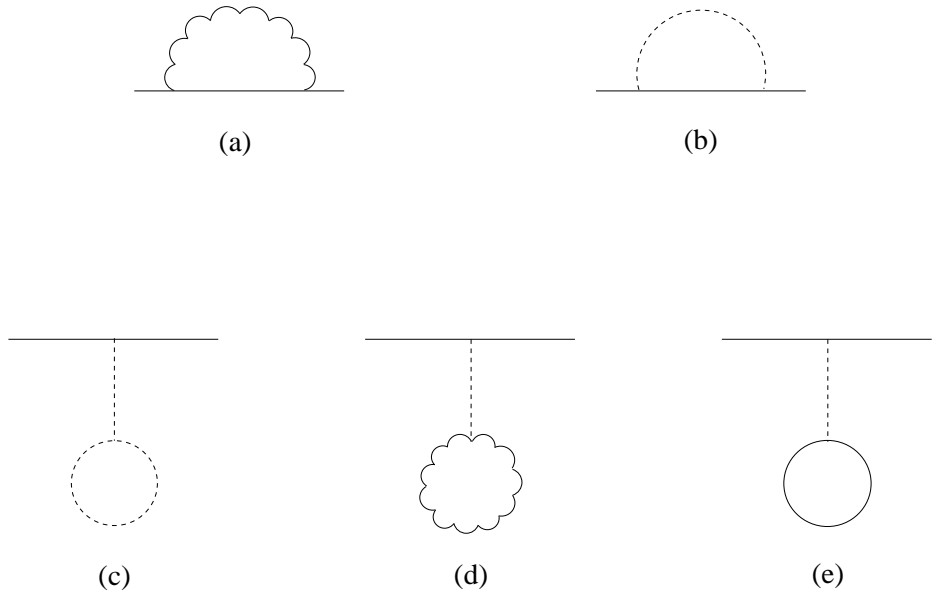,width=0.9\textwidth}}
\vspace{1cm}  
\caption{One loop corrections to the fermion mass matrix, where solid lines
represent fermions, wavy lines dual gauge bosons and dotted lines dual
colour Higgs fields.}
\label{oneloop}
\end{figure}

Our fermions carry in fact weak isospin and in the case of quarks 
also colour, so that in principle there will be loop corrections
coming from colour gluon and electroweak boson loops.  However, as far as
the CKM matrix, or the lower generation fermion masses, are concerned,
only those diagrams which rotate the mass matrix with respect to dual
colour (i.e.\ the generation index) will matter.  Since neither the colour 
gluons nor the gauge and Higgs bosons in the electroweak sector carry
dual colour, they cannot rotate the generation index, and hence will leave
both factors (\ref{fermassmat}) of the mass matrix intact, affecting at most
its normalization.  As we shall see, there are other reasons why we cannot
in any case concern ourselves with the normalization of the mass matrix. 
There is thus no point for us to consider gluon and electroweak boson
loops any further.  There remain then only those diagrams with dual gauge 
and Higgs boson loops, of which all those of 1-loop order are listed in 
Figure \ref{oneloop}.

Let us first write down the explicit expressions for the corrections to the 
fermion mass matrix arising from the diagrams in Figure \ref{oneloop}.  This 
has been done already in a general framework by Weinberg \cite{Weinberg} 
who expressed the answer as a sum of five
terms:
\begin{equation}
\delta m = - \Sigma^{(A1)} - \Sigma^{(A \phi)} - \Sigma^{(AT)}
   -\Sigma^{(\phi 1)} - \Sigma^{(T1)},
\label{deltam}
\end{equation}
where:
\begin{eqnarray}
\Sigma^{(A1)}\!\!&\!\!=\!\!&\!\!\frac{1}{16 \pi^2} \sum_N \int_0^1 dx 
   \{-2m \bar{t}_N (1-x) + 4 \gamma_4 \bar{t}_N \gamma_4 m\} \ln \left(\mu_N^2
   + \frac{m^2 x^2}{1-x} \right) \bar{t}_N, \nonumber \\
\Sigma^{(A \phi)}\!\!&\!\!=\!\!&\!\!\frac{1}{16 \pi^2} \sum_N \frac{1}{\mu_N^2}
   \int_0^1 dx \{(1-x) m [\gamma_4 m, \bar{t}_N] \gamma_4
   + \gamma_4 [\gamma_4 m, \bar{t}_N] m \} \nonumber \\
   \!\!&\!\!  \!\!&\!\!\times \left\{ \ln \left( \frac{m^2 x^2}{1-x} \right)
   - \ln \left( \mu_N^2 + \frac{m^2 x^2}{1-x} \right) \right\}
   \gamma_4 [\gamma_4 m, \bar{t}_N], \nonumber \\
\Sigma^{(AT)}\!\!&\!\!=\!\!&\!\!\frac{1}{32 \pi^2} \sum_N \gamma_4 [\bar{t}_N,
   [\bar{t}_N, \gamma_4 m]] \ln \mu_N^2, \nonumber \\
\Sigma^{(\phi 1)}\!\!&\!\!=\!\!&\!\! - \frac{1}{16 \pi^2} 
\sum_K\!\int_0^1\!\!dx
   \{-(1-x) m \gamma_4 \bar{\Gamma}_K \gamma_4 + \bar{\Gamma}_K m\}
   \ln\{m^2 x^2 + M_K^2 (1-x)\} \bar{\Gamma}_K, \nonumber \\
\Sigma^{(T1)}\!\!&\!\!=\!\!&\!\!\frac{1}{32 \pi^2} 
\Gamma_i M^{-2}_{\ \ ij}
   \left\{\!f_{klj}(M^2 \ln M^2)_{kl}\!\!-\!\!16 {\rm Tr} (m^3 \ln m 
   \Gamma_j)\!\!+\!\!6
   \sum_N (\bar{\theta}_N^2 \lambda)_j \mu_N^2 \ln \mu_N^2
   \right\}.
   \nonumber \\
&&
\label{Sigmas}
\end{eqnarray}
As they are written, these formulae depend on the energy scale, the 
significance of which will be elucidated later.

The above formulae depend also on the following quantities, the explicit 
forms for which have yet to be specified: the fermion mass matrix $m$,
the eigenvalues $\mu_N, N = 0, ..., 8$ of the dual gauge boson mass matrix,
the Higgs boson mass matrix $M_{ij}$ and its eigenvalues $M_K, K = 1, ..., 9$,
the fermion couplings to the dual gauge bosons $\bar{t}_N$ and to the Higgs
bosons $\Gamma_i$ or $\bar{\Gamma}_K$, and then the Higgs bosons' couplings
to themselves $f_{klj}$ and to the dual gauge bosons $(\bar{\theta}_N^2
\lambda)_j$.  We proceed to do so now.

The fermion mass matrix Weinberg defined somewhat differently from that
given above in (\ref{fermassmat}).  Writing the Yukawa coupling in terms 
of the full fermion field $\psi$, thus:
\begin{equation}
\sum_{[b]} Y_{[b]} \sum_{(a)} \bar{\psi}^{\tilde{a}} \phi^{(a)}_{\tilde{a}} 
   \psi^{[b]} + {\rm h.c.}, 
\label{Yukawaw}
\end{equation}
instead of the left- and right-handed components as in (\ref{Yukawa}), one
obtains a mass matrix of the following form:
\begin{equation}
m_{W'} = m \frac{1}{2} (1 + \gamma_5) + m^{\dagger} \frac{1}{2} (1 - \gamma_5),
\label{fermassmatw1}
\end{equation}
containing factors in $\gamma_5$.  However, since the labels on the 
right-handed components have actually no physical significance, one can 
relabel them such as to make $m$ hermitian and hence obtain the mass matrix 
in the form:
\begin{equation}
m_W = \rho \zeta \left( \begin{array}{c} x \\ y \\ z \end{array} \right)
   (x, y, z),
\label{fermassmatw}
\end{equation}
which has no $\gamma_5$ in it, and is essentially just the square root of
$m m^{\dagger}$ in terms of the $m$ previously defined.  We notice that $m_W$
remains a factorized matrix, which property is crucial for our discussion
later.  In the calculations which follow, when no confusion is likely to 
occur, we shall drop the subscript $W$ from the Weinberg mass matrix.

Next, the mass matrix for the dual gauge bosons has already been worked
out in \cite{Chantsou}.  This $9 \times 9$ matrix is diagonal for
$\alpha = N = 1, 2, 4, 5, 6, 7$, as labelled by the Gell-Mann matrices
$\lambda_\alpha$ of $\widetilde{SU}(3)$, with eigenvalues:
\begin{eqnarray}
\mu_1, \mu_2 & = & \frac{\tilde{g}_3 \zeta}{2} \sqrt{x^2 + y^2}, \\ \nonumber
\mu_4, \mu_5 & = & \frac{\tilde{g}_3 \zeta}{2} \sqrt{z^2 + x^2}, \\ \nonumber
\mu_6, \mu_7 & = & \frac{\tilde{g}_3 \zeta}{2} \sqrt{y^2 + z^2}.
\label{muNdiag}
\end{eqnarray}
The remaining $3 \times 3$ nondiagonal block in $\mu^2$ as labelled by 
$\lambda_3$, $\lambda_8$ and $\lambda_0 = \frac{2}{3}I$ reads as:
\begin{equation}
\left( \begin{array}{ccc}
   \frac{\tilde{g}_3^2}{4} \zeta^2 (x^2+y^2) & \frac{\tilde{g}_3^2}{4 \sqrt{3}}
   \zeta^2 (x^2-y^2) & - \frac{\tilde{g}_1 \tilde{g}_3}{3} \zeta^2 (x^2-y^2) \\
   \frac{\tilde{g}_3^2}{4 \sqrt{3}} \zeta^2 (x^2-y^2) & \frac{\tilde{g}_3^2}{12}
   \zeta^2 (x^2+y^2+4z^2) & - \frac{\tilde{g}_1 \tilde{g}_3}{3\sqrt{3}}
   \zeta^2 (x^2+y^2-2z^2) \\
   - \frac{\tilde{g}_1 \tilde{g}_3}{3} \zeta^2 (x^2-y^2) & - \frac{\tilde{g}_1
   \tilde{g}_3}{3 \sqrt{3}} \zeta^2 (x^2+y^2-2z^2) & \frac{4 \tilde{g}_1^2}{9}
   \zeta^2 (x^2+y^2+z^2) \end{array} \right),
\label{munondiag}
\end{equation}
the eigenvalues of which we label by $\mu_N^2, N = 3, 8, 0$ with eigenvectors
$C_{\alpha N}$ such that:
\begin{equation}
\sum_{\alpha, \beta = 3, 8, 0} C_{\alpha N'} \mu_{\alpha \beta}^2 C_{\beta N} 
   = \mu_N^2 \delta_{N N'}.
\label{CalphaN}
\end{equation}
The diagonalization of this matrix we shall perform only with explicit
values for the parameters. 

The Higgs fields $\phi^{(a)}_{\tilde{a}}$ represent 9 complex degrees of
freedom, which we write, following Weinberg's convention, as 18 real fields,
thus:
\begin{equation}
\phi^{(a)}_{\tilde{a}} = \phi^{(a)}_{\tilde{a} 1} + i \phi^{(a)}_{\tilde{a} 2}.
\label{phireal}
\end{equation}
From these and the potential $V[\phi]$ in (\ref{Higgspot}), the Higgs
boson mass matrix is given by:
\begin{equation}
M^2 = \left[ \frac{\partial^2 V}{\partial \phi^{(a)}_{\tilde{a} r}
   \partial \phi^{(b)}_{\tilde{b} s}} \right]_{vacuum},
\label{Higmassmat}
\end{equation}
which breaks up into 8 diagonal blocks as follows.  First, there is a 
$3 \times 3$ block labelled by $\phi^{(1)}_{1,1}, \phi^{(2)}_{2,1},
\phi^{(3)}_{3,1}$:
\begin{equation}
8 \lambda \zeta^2 \left( \begin{array}{ccc} x^2 & xy & xz \\
                          yx & y^2 & yz \\
                          zx & zy & z^2 
       \end{array} \right).
\label{Mblock1}
\end{equation}
Second, there is a $2 \times 2$ block labelled by $\phi^{(3)}_{2,1},
\phi^{(2)}_{3,1}$:
\begin{equation}
4 \kappa \zeta^2 \left( \begin{array}{cc} y^2 & yz \\
                         zy & z^2 \end{array} \right).
\label{Mblock2}
\end{equation}
Third, there is a $2 \times 2$ block labelled by $\phi^{(3)}_{2,2},
\phi^{(2)}_{3,2}$:
\begin{equation}
4 \kappa \zeta^2 \left( \begin{array}{cc} y^2 & - yz \\
                         - zy & z^2 \end{array} \right).
\label{Mblock3}
\end{equation}
Fourth, there is a $2 \times 2$ block labelled by $\phi^{(1)}_{3,1},
\phi^{(3)}_{1,1}$:
\begin{equation}
4 \kappa \zeta^2 \left( \begin{array}{cc} z^2 & zx \\
                         xz & x^2 \end{array} \right).
\label{Mblock4}
\end{equation}
Fifth, there is a $2 \times 2$ block labelled by $\phi^{(1)}_{3,2},
\phi^{(3)}_{1,2}$:
\begin{equation}
4 \kappa \zeta^2 \left( \begin{array}{cc} z^2 & - zx \\
                         - xz & x^2 \end{array} \right).
\label{Mblock5}
\end{equation}
Sixth, there is a $2 \times 2$ block labelled by $\phi^{(2)}_{1,1},
\phi^{(1)}_{2,1}$:
\begin{equation}
4 \kappa \zeta^2 \left( \begin{array}{cc} x^2 & xy \\
                         yx & y^2 \end{array} \right).
\label{Mblock6}
\end{equation}
Seventh, there is a $2 \times 2$ block labelled by $\phi^{(2)}_{1,2},
\phi^{(1)}_{2,2}$:
\begin{equation}
4 \kappa \zeta^2 \left( \begin{array}{cc} x^2 & - xy \\
                         - yx & y^2 \end{array} \right).
\label{Mblock7}
\end{equation}   
Finally, there is a $3 \times 3$ block labelled by $\phi^{(1)}_{1,2},
\phi^{(2)}_{2,2}, \phi^{(3)}_{3,2}$ the entries of which are all zero.
All the first seven blocks are of rank 1 and have each only one nonzero
eigenvalue, giving thus for $M^2$ altogether 11 zero modes, 9 of which,
namely one each from each block except the first and last and all 3
from the last, are eaten up by the dual gauge bosons, leaving 2 from the
first block.  These 2 remaining zero modes come from an ``accidental
symmetry'' of the vacuum, not from a symmetry of the potential (\ref{Higgspot}),
as explained in \cite{Chantsou}, and are therefore not eaten up by the 
gauge bosons.  We are then left with 7 Higgs bosons coming one from 
each of the first seven blocks which we label in that order with their 
eigenvalues and eigenvectors each in its own block as follows:
\begin{eqnarray}
K = 1: & 8 \lambda \zeta^2 (x^2+y^2+z^2) & (x, y, z), \nonumber \\
K = 2: & 4 \kappa \zeta^2 (y^2 + z^2)    & (y, z), \nonumber \\
K = 3: & 4 \kappa \zeta^2 (y^2 + z^2)    & (y, -z), \nonumber \\
K = 4: & 4 \kappa \zeta^2 (z^2 + x^2)    & (z, x), \nonumber \\
K = 5: & 4 \kappa \zeta^2 (z^2 + x^2)    & (z, -x), \nonumber \\
K = 6: & 4 \kappa \zeta^2 (x^2 + y^2)    & (x, y), \nonumber \\
K = 7: & 4 \kappa \zeta^2 (x^2 + y^2)    & (x, -y),
\label{MK1to7}
\end{eqnarray}
while the two remaining zero (pseudo-Goldstone) modes coming from the 
first block will be assigned the following eigenvectors in the original 
basis of that block:
\begin{equation}
|v_8 \rangle = -\beta \left( \begin{array}{c} y-z \\ z-x \\ 
   x-y \end{array} \right);
|v_9 \rangle = \beta \left( \begin{array}{c} 1-x(x+y+z) \\ 1-y(x+y+z) \\ 
   1-z(x+y+z) \end{array} \right),
\label{vK8to9}
\end{equation}
with
\begin{equation}
\beta^{-2} = 3 - (x+y+z)^2.
\label{beta}
\end{equation}

Next, the couplings of the dual gauge bosons to the fermions are in
Weinberg's convention:
\begin{eqnarray}
\bar{t}_N & = & t_N, \ \ \  N = 1, 2, 4, 5, 6, 7, \nonumber \\
\bar{t}_N & = & C_{3N} t_3 + C_{8N} t_8 + C_{0N} t_0, \ \ \  N = 3, 8, 0,
\label{tbarN}
\end{eqnarray}
where:
\begin{eqnarray}
t_\alpha & = & - \frac{\tilde{g}_3}{2} \lambda_\alpha \frac{1}{2} (1-\gamma_5),
   \ \ \  \alpha = 1, ..., 8, \nonumber \\
t_0 & = & \frac{2 \tilde{g}_1}{3} \frac{1}{2} (1-\gamma_5).
\label{talpha}
\end{eqnarray}
The coefficients $C_{3N}, C_{8N}, C_{0N}$ are as defined in (\ref{CalphaN}).

The couplings of the Higgs bosons to the fermions as extracted from the
Yukawa couplings (\ref{Yukawaw}), expressed in terms of the real Higgs
fields (\ref{phireal}), and rotated to the basis where $m$ is hermitian
or where $m_W$ has no $\gamma_5$, are:
\begin{eqnarray}
\Gamma^{(a)}_{11} & = & \frac{1}{2} (1 + \gamma_5) \rho \left(
   \begin{array}{ccc} x & y & z \\ 0 & 0 & 0 \\ 0 & 0 & 0 \end{array} \right)
     + \frac{1}{2} (1 - \gamma_5) \rho \left(
   \begin{array}{ccc} x & 0 & 0 \\ y & 0 & 0 \\ z & 0 & 0 \end{array} \right),
   \nonumber \\
\Gamma^{(a)}_{12} & = & \frac{i}{2} (1 + \gamma_5) \rho \left(
   \begin{array}{ccc} x & y & z \\ 0 & 0 & 0 \\ 0 & 0 & 0 \end{array} \right)
     - \frac{i}{2} (1 - \gamma_5) \rho \left(
   \begin{array}{ccc} x & 0 & 0 \\ y & 0 & 0 \\ z & 0 & 0 \end{array} \right),
   \nonumber \\
\Gamma^{(a)}_{21} & = & \frac{1}{2} (1 + \gamma_5) \rho \left(
   \begin{array}{ccc} 0 & 0 & 0 \\ x & y & z \\ 0 & 0 & 0 \end{array} \right)
     + \frac{1}{2} (1 - \gamma_5) \rho \left(
   \begin{array}{ccc} 0 & x & 0 \\ 0 & y & 0 \\ 0 & z & 0 \end{array} \right),
   \nonumber \\
\Gamma^{(a)}_{22} & = & \frac{i}{2} (1 + \gamma_5) \rho \left(
   \begin{array}{ccc} 0 & 0 & 0 \\ x & y & z \\ 0 & 0 & 0 \end{array} \right)
     - \frac{i}{2} (1 - \gamma_5) \rho \left(
   \begin{array}{ccc} 0 & x & 0 \\ 0 & y & 0 \\ 0 & z & 0 \end{array} \right),
   \nonumber \\
\Gamma^{(a)}_{31} & = & \frac{1}{2} (1 + \gamma_5) \rho \left(
   \begin{array}{ccc} 0 & 0 & 0 \\ 0 & 0 & 0 \\ x & y & z \end{array} \right)
     + \frac{1}{2} (1 - \gamma_5) \rho \left(
   \begin{array}{ccc} 0 & 0 & x \\ 0 & 0 & y \\ 0 & 0 & z \end{array} \right),
   \nonumber \\
\Gamma^{(a)}_{32} & = & \frac{i}{2} (1 + \gamma_5) \rho \left(
   \begin{array}{ccc} 0 & 0 & 0 \\ 0 & 0 & 0 \\ x & y & z \end{array} \right)
     - \frac{i}{2} (1 - \gamma_5) \rho \left(
   \begin{array}{ccc} 0 & 0 & x \\ 0 & 0 & y \\ 0 & 0 & z \end{array} \right),
\label{Gamma}
\end{eqnarray}
which are independent of the superscript $(a)$.  Notice that the three
indices $(a) = 1, 2, 3; \tilde{a} = 1, 2, 3; r = 1, 2$ here are to play
together the role of the index $i = 1, ..., 18$ in the formula for 
$\Sigma^{T1}$ in (\ref{Sigmas}).  Alternatively, when expressed in terms 
of the basis formed by the eigenstates of the Higgs mass matrix $M$ as 
listed in (\ref{MK1to7}) and (\ref{vK8to9}), we have the same couplings 
in the form to be used in $\Sigma^{\phi 1}$ of (\ref{Sigmas}):
\begin{equation}
\bar{\Gamma}_K = \bar{\gamma}_K \frac{1}{2} (1 + \gamma_5)
   + \bar{\gamma}^{\dagger}_K \frac{1}{2} (1 - \gamma_5),
\label{Gammabar}
\end{equation}
where:
\begin{equation}
\bar{\gamma}_K = \rho |v_K \rangle \langle v_1|,
\label{gammabar}
\end{equation}
and:
\begin{eqnarray}
|v_1 \rangle & = & \left( \begin{array}{c} x \\ y \\ z \end{array} \right),
   \nonumber \\
|v_2 \rangle & = & \frac{1}{\sqrt{y^2 + z^2}}
   \left( \begin{array}{c} 0 \\ y \\ z \end{array} \right), \nonumber \\
|v_3 \rangle & = & \frac{i}{\sqrt{y^2 + z^2}}
   \left( \begin{array}{c} 0 \\ y \\-z \end{array} \right), \nonumber \\
|v_4 \rangle & = & \frac{1}{\sqrt{z^2 + x^2}}
   \left( \begin{array}{c} x \\ 0 \\ z \end{array} \right), \nonumber \\
|v_5 \rangle & = & \frac{i}{\sqrt{z^2 + x^2}}
   \left( \begin{array}{c} -x\\ 0 \\ z \end{array} \right), \nonumber \\
|v_6 \rangle & = & \frac{1}{\sqrt{x^2 + y^2}}
   \left( \begin{array}{c} x \\ y \\ 0 \end{array} \right), \nonumber \\
|v_7 \rangle & = & \frac{i}{\sqrt{x^2 + y^2}}
   \left( \begin{array}{c} x \\-y \\ 0 \end{array} \right),
\label{vK1to7}
\end{eqnarray}
while $|v_8 \rangle$ and $|v_9 \rangle$ are already given in (\ref{vK8to9}).

There remains then for us to specify only the couplings 
$(\bar{\theta}_N^2 \lambda)_j$
of the gauge fields to the Higgs fields and the couplings $f_{klj}$ of the
Higgs fields to themselves, both occurring in the tadpole term $\Sigma^{T1}$
in (\ref{Sigmas}).  The calculations for these are somewhat tedious, 
especially $f_{klj}$ which has altogether $18 \times 18 \times 18$ entries, 
most of which are zero.  Since the calculation is straightforward and their 
actual values will not be needed for our calculation later, here we shall 
give only those results which are relevant for our considerations.  For 
instance, one does not need to include in the sum over $K$ the zero modes 
(\ref{vK8to9}) \cite{Weinberg}.  For the rest, it is easier to state the 
result in terms of the basis formed by the eigenstates (\ref{MK1to7}) of 
the Higgs mass matrix than in terms of the original basis labelled by the 
three indices $(a) = 1, 2, 3, \tilde{a} = 1, 2, 3, r = 1, 2$ corresponding 
together to the index $i$ in Weinberg's formulae.  In that case, the couplings 
$(\bar{\theta}_N^2 \lambda)_K$ for $K = 2, ..., 7$ all vanish, leaving only:
\begin{equation}
(\bar{\theta}_N^2 \lambda)_{K=1} = - \zeta (D_{N1}^2 x^2 + D_{N2}^2 y^2 
   + D_{N3}^2 z^2),
\label{thetabar}
\end{equation}
with
\begin{eqnarray}
D_{N1}^2 = \frac{\tilde{g}_3^2}{4}, D_{N2}^2 = \frac{\tilde{g}_3^2}{4},
   D_{N3}^2 = 0; \ \ \  N & = & 1, 2; \nonumber \\
D_{N1}^2 = \frac{\tilde{g}_3^2}{4}, D_{N2}^2 = 0, D_{N3}^2 =
   \frac{\tilde{g}_3^2}{4}; \ \ \  N & = & 4, 5; \nonumber \\
D_{N1}^2 = 0, D_{N2}^2 = \frac{\tilde{g}_3^2}{4}, D_{N3}^2 =
   \frac{\tilde{g}_3^2}{4}; \ \ \  N & = & 6, 7,
\label{Dnot380}
\end{eqnarray}
and, for $N = 3, 8, 0$:
\begin{eqnarray}
D_{N1}^2 & = & \left( C_{3N} \frac{\tilde{g}_3}{2} + C_{8N} \frac{\tilde{g}_3}
   {2\sqrt{3}} - C_{0N} \frac{2 \tilde{g}_1}{3} \right) ^2 \nonumber \\
D_{N2}^2 & = & \left(-C_{3N} \frac{\tilde{g}_3}{2} + C_{8N} \frac{\tilde{g}_3}
   {2\sqrt{3}} - C_{0N} \frac{2 \tilde{g}_1}{3} \right) ^2 \nonumber \\
D_{N3}^2 & = & \left(-C_{8N} \frac{\tilde{g}_3}{\sqrt{3}}
   - C_{0N} \frac{2 \tilde{g}_1}{3} \right) ^2.
\label{D380}
\end{eqnarray}
Finally, in the same basis of eigenstates, of the Higgs fields self-couplings
$\bar{f}_{IJK}$ we need only those with $I = J$ and these are found to
vanish except when $K = 1$, where:
\begin{eqnarray}
\bar{f}_{111} & = & 24 \lambda \zeta, \nonumber \\
\bar{f}_{221} & = & 8 \lambda \zeta + 8 \kappa\zeta (y^2 + z^2), \nonumber \\
\bar{f}_{331} & = & 8 \lambda \zeta + 16 \kappa\zeta \left(\frac{y^2
 z^2}{y^2 + z^2}  \right) \nonumber \\
\bar{f}_{441} & = & 8 \lambda \zeta + 8 \kappa\zeta (z^2 + x^2), \nonumber \\
\bar{f}_{551} & = & 8 \lambda \zeta + 16 \kappa\zeta \left(\frac{z^2 x^2}{z^2 + x^2}
   \right) \nonumber \\
\bar{f}_{661} & = & 8 \lambda \zeta + 8 \kappa\zeta (x^2 + y^2), \nonumber \\
\bar{f}_{771} & = & 8 \lambda \zeta + 16 \kappa\zeta 
   \left(\frac{x^2 y^2}{x^2 + y^2} \right).
\label{fbar}
\end{eqnarray}
With these, the specification of quantities entering in the expressions
for the 1-loop diagrams in (\ref{Sigmas}) is complete.

\setcounter{equation}{0}

\section{The Relevant Terms}

Although the 1-loop diagrams detailed in the preceding section were all 
referred to formally as corrections, they need not all be small.  In
particular, the coupling of the dual gluon is given in terms of that of 
the usual colour gluon by the Dirac quantization condition \cite{Chantsoua}:
\begin{equation}
g \tilde{g} = 4 \pi,
\label{Diraccond}
\end{equation}
which means that for the usual colour coupling $g$ having the observed
value of around 1.18 at the $Z$ mass, the dual colour coupling $\tilde{g}$ 
is of order 10.  Thus, loop diagrams such as Figure \ref{oneloop}(a) 
and (d), in which the integrated momenta need not be low so that the propagator 
suppression by a high dual gluon mass is inoperative, can in fact take 
on very large values.  They cannot then be treated perturbatively.

However, not all the diagrams in Figure \ref{oneloop} rotate the fermion
mass matrix with respect to dual colour, which rotation is needed to give 
a nontrivial CKM matrix and nonzero masses to the two lower generations.  
Indeed, as we shall see, it turns out that the large contributions from 
Figure \ref{oneloop} will all affect only the normalisation of the 
mass matrix but not its orientation in dual colour, so that as far as 
the effects we seek are concerned, there are only small corrections to be 
considered.  This is very fortunate, for otherwise one would not be able to
calculate the lower generation quark masses and the CKM matrix perturbatively
as we set out to do.

To see that this fortunate situation does indeed arise, consider first the 
last term in $\Sigma^{T1}$ in (\ref{Sigmas}) which represents the dual gauge 
boson tadpole diagram exhibited in Figure \ref{oneloop}(d).  This term is huge, 
being proportional to $\tilde{g}_3^2 \mu_N^2 \ln \mu_N^2$, where $\tilde{g}_3$, 
as already mentioned above, is of order 10, and the dual gauge bosons, in 
order for their exchanges not to violate the very stringent experimental 
bounds on flavour-changing neutral current decays, have to have masses 
$\mu_N$ of the order of 100 TeV \cite{Cahnrari,Bordesetal}.  However, in 
terms of the basis of Higgs mass eigenstates, this term appears as:
\begin{equation}
\frac{3}{16 \pi^2} \sum_K \bar{\Gamma}_K M_K^{-2} \sum_N 
   (\bar{\theta}_N^2 \lambda)_K \mu_N^2 \ln \mu_N^2.
\label{gtadpole}
\end{equation}
Now, according to our previous result stated in (\ref{thetabar}), only the
Higgs state $K = 1$ has nonvanishing coupling to the dual gauge bosons 
so that the sum over $K$ in (\ref{gtadpole}) has only the $K=1$ term.
By (\ref{Gammabar})-(\ref{vK1to7}), 
however, $\bar{\Gamma}_1$ is itself proportional to the tree-level mass matrix
$m_W$ in (\ref{fermassmatw}), so that the whole diagram has the effect only
of changing the normalization of the tree-level mass matrix as anticipated.

A similar conclusion is reached for the other terms $\Sigma^{A1}$,
$\Sigma^{A \phi}$ and $\Sigma^{AT}$ in (\ref{Sigmas}), coming from the dual 
gauge boson loop. In contrast to the dual gauge boson tadpole studied in the 
above paragraph, these terms do rotate the fermion mass matrix but do so only
through the mass matrix $m$ itself on which these terms depend.  Suppose
then we expand these expressions in powers of $m$, then the leading term
of order $m$ will be just a scalar times the original tree-level mass 
matrix and can therefore only affect its normalization, not its orientation.  
The other terms in the expansion which rotate the mass matrix will be 
of order $m^2/\mu_N^2$ times the mass matrix and hence much smaller, in 
fact even negligible, as we shall see later.

The fact that the normalization of the fermion mass matrix is affected by
large loop corrections means of course that we cannot hope to calculate
its value perturbatively but have to treat it as a parameter to be
determined experimentally.  It means in particular that the one nonzero 
eigenvalue of $m$ corresponding to the mass of the highest generation which
started at tree-level as $\rho \zeta$ can now no longer be given in terms
of $\rho$, the Yukawa coupling, and $\zeta$ the vacuum expectation value
of the Higgs fields, but has to be treated as a separate parameter, say $m_T$
(where $T$ labels the type of fermions under consideration, namely $U$ or 
$D$ for quarks and $L$ for leptons), thus reducing the predictive power 
of the present calculation.  On the other hand, since the normalization 
cannot be predicted in any case, there is now no sense in calculating 
those diagrams which affect the normalization only.  Hence, as mentioned 
already in the beginning of this section, one may safely ignore all those 
diagrams with loops of the ordinary colour gluons and the usual electroweak 
Higgs bosons.  

A similar conclusion applies also to the dual Higgs tadpole diagram in
Figure \ref{oneloop}(c) representing the first term in $\Sigma^{T1}$ of 
(\ref{Sigmas}).  In the diagonal basis for the Higgs mass matrix $M$, 
this term appears as:
\begin{equation}
\frac{1}{32 \pi^2} \sum_K \sum_L \bar{\Gamma}_K M_K^2 \bar{f}_{LLK}
   (M^2 \ln M^2)_L,
\label{Htadpole}
\end{equation}
where since, according to (\ref{fbar}), $\bar{f}_{LLK}$ vanish except for 
$K = 1$, there remains only one term in the sum proportional to the matrix 
$\bar{\Gamma}_1$, which is itself proportional to the tree-level mass
matrix $m$.  It will therefore affect only the normalization of $m$, not 
its orientation, and thus, by the logic above, can also be ignored.  Given 
that in Figure \ref{oneloop}, only the diagram (c) depends on the Higgs 
self-couplings $f_{ijk}$, this means that we can henceforth eliminate this coupling 
from our considerations.

The remaining terms in Figure \ref{oneloop}, namely the terms of order
$m^2/\mu_N^2$ or higher in $\Sigma^{A1}$, $\Sigma^{A \Phi}$, and $\Sigma^{AT}$,
the term $\Sigma^{\phi 1}$, and the second term in $\Sigma^{T1}$ corresponding 
to the fermion-loop tadpole of Figure \ref{oneloop}(e), the explicit expressions
for which are given in (\ref{Sigmas}), all rotate the mass matrix.  However, 
as explained in ref. \cite{Chantsou}, they will leave the mass matrix in a 
factorized form with only the left-handed factor rotated.  In other words, the
correction $\Sigma$ from these terms can each be written in the form:
\begin{equation}
\Sigma = m_T \left( \begin{array}{c} x_{1+} \\ y_{1+} \\ z_{1+}
   \end{array} \right) (x, y, z) \frac{1}{2} (1 + \gamma_5)
   + m_T \left( \begin{array}{c} x \\ y \\ z \end{array} \right)
   (x_{1-}, y_{1-}, z_{1-}) \frac{1}{2} (1 - \gamma_5).
\label{delsigma}
\end{equation}
Added to the tree-level mass matrix and symmetrized with respect to left and right this gives:
\begin{equation}
\Sigma = m_T \left( \begin{array}{c} x' \\ y' \\ z' \end{array} \right)
   (x, y, z) \frac{1}{2} (1 + \gamma_5)
   + m_T \left( \begin{array}{c} x \\ y \\ z \end{array} \right)
   (x', y', z') \frac{1}{2} (1 - \gamma_5),
\label{fermassmatp}
\end{equation}
where $x' = x - \frac{1}{2} x_{1+} - \frac{1}{2} x_{1-}$ etc.  Again, as in the tree-level mass 
matrix, the $\gamma_5$ terms in (\ref{fermassmatp}) can be rotated away
by redefining the right-handed fermions to give:
\begin{equation}
m'_W = \frac{m_T}{\zeta'^2} \left( \begin{array}{c} x' \\ y' \\ z' 
   \end{array} \right) (x', y', z').
\label{fermassmatpW}
\end{equation}
Hence, in order to specify the effect of the loop corrections, we need 
only give for each the corresponding $x_1 = \frac{1}{2} x_{1+} +
\frac{1}{2} x_{1-}$ etc.

The results of our calculation are as follows.  From the sum of $\Sigma^{A1}$
and $\Sigma^{AT}$ in (\ref{Sigmas}), we obtain:
\begin{equation}
\left( \begin{array}{c} x_1 \\ y_1 \\ z_1 \end{array} \right)
   = -\frac{1}{8 \pi^2} \sum_N \bar{T}_N f_N(m^2) \bar{T}_N
   \left( \begin{array}{c} x \\ y \\ z \end{array} \right),
\label{x1A1AT}
\end{equation}
where $\bar{T}_N$ are just the dual gluon couplings $\bar{t}_N$ listed in
(\ref{tbarN}) without the $\frac{1}{2} (1 - \gamma_5)$ factor:
\begin{equation}
\bar{t}_N = \bar{T}_N \frac{1}{2} (1 - \gamma_5),
\label{TbarN}
\end{equation}
and $f_N(m^2)$ is the integral:
\begin{equation}
f_N(m^2) = \int_0^1 dx (1-x) \ln \left[1 + \frac{(m^2/\mu_N^2) x^2}
   {1-x} \right].
\label{fN}
\end{equation}
From $\Sigma^{A \phi}$ (\ref{Sigmas}), we have:
\begin{equation}
\left( \begin{array}{c} x_1 \\ y_1 \\ z_1 \end{array} \right)
   = \frac{1}{16 \pi^2} \sum_N \frac{1}{\mu_N^2} \left[-m^2 \bar{T}_N
   g_N^-(m^2) \bar{T}_N + \bar{T}_N m^2 g_N^+(m^2) \bar{T}_N \right]
   \left( \begin{array}{c} x \\ y \\ z \end{array} \right),
\label{x1Aphi}
\end{equation}
with:
\begin{equation}
g_N^{\pm}(m^2) = \int_0^1 dx (1 \pm x) \left[ \ln \left(\frac{m^2 x^2}{1-x}
   \right) - \ln \left( \mu_N^2 + \frac{m^2 x^2}{1-x} \right) \right].
\label{g+-N}
\end{equation}
From $\Sigma^{\phi 1}$ in (\ref{Sigmas}) we obtain:
\begin{equation}
\left( \begin{array}{c} x_1 \\ y_1 \\ z_1 \end{array} \right)
   = - \frac{1}{16 \pi^2} \left[\sum_K A_K |v_1 \rangle 
   + \sum_K B_K |v_K \rangle \right],
\label{x1phi1}
\end{equation}
where $|v_K \rangle$ are as listed in (\ref{vK1to7}) and (\ref{vK8to9}). 
The first term with:
\begin{equation}
A_K = -  \rho^2 \langle v_K| F_K(m^2) |v_K \rangle
\label{AK}
\end{equation}
need not bother us, being proportional to $|v_1 \rangle$ which is the same 
as for the tree-level mass matrix.  For the second term, we have:
\begin{equation}
B_K = - \rho^2 \langle v_1| F_K(m^2) |v_1 \rangle \langle v_K|v_1 \rangle
   + 2 \rho^2 \langle v_1| G_K(m^2) |v_1 \rangle \langle v_1|v_K \rangle,
\label{BK}
\end{equation}
with
\begin{equation}
F_K(m^2) = \int_0^1 dx (1-x) \ln [m^2 x^2 +M_K^2(1-x)],
\label{FK}
\end{equation}
and
\begin{equation}
G_K(m^2) = \int_0^1 dx \ln [m^2 x^2 + M_K^2 (1-x)].
\label{GK}
\end{equation}
Lastly, from the fermion-loop tadpole term of Figure \ref{oneloop}(e) as
given in the second term of $\Sigma^{T1}$ in (\ref{Sigmas}), we obtain:
\begin{equation}
\left( \begin{array}{c} x_1 \\ y_1 \\ z_1 \end{array} \right)
   = - \frac{1}{\pi^2} \frac{\rho^2}{m_T} \sum_K M_K^{-2} {\rm Tr}
   \{m^3 \ln m \bar{\gamma}_K \} |v_K \rangle,
\label{x1T1f}
\end{equation}
with $\bar{\gamma}_K$ as given in (\ref{gammabar}).  The sum here is to
be taken only over the Higgs bosons with nonzero masses, namely over only
$K = 1, ..., 7$ with $M_K$ given in (\ref{MK1to7}).

Although the terms listed in the preceding paragraph all rotate the mass
matrix and hence could contribute to the present calculation of the CKM 
matrix and lower generation masses, they are widely different in size.
Thus, the terms (\ref{x1A1AT}) and (\ref{x1Aphi}) are both of order
$m^2/\mu_N^2$ where $m$ is about 176 GeV for $U$-type quarks and  
4.3 GeV for $D$-type quarks, while, as already mentioned before, the dual 
gauge bosons are bounded by experiment to have masses larger than 100 TeV,
a bound that we shall be able to check later within the present framework
for consistency.  That being the case, the corrections due to these two 
terms are only of the order of $10^{-6}$ or less and are thus seen to be 
entirely negligible for calculating the lower generation quark masses or 
the CKM matrix to the present experimental accuracy.  A similar conclusion 
is reached also for the term (\ref{x1T1f}) which is of order 
$(m^2 \ln m)/M_K^2$, 
with Higgs boson masses $M_K$ being estimated to be of order some tens of 
TeV, again an estimate that we shall be able to check for consistency
within the present framework.  Hence, the end result of our analysis is 
that of all the 1-loop corrections we have evaluated, only the term 
(\ref{x1phi1}) from the Higgs boson loop in Figure \ref{oneloop}(b) 
is potentially large enough to give the right orders of magnitude for 
the lower generation quark masses or for the off-diagonal CKM matrix 
elements, and it is therefore to this term that we shall now direct our 
attention.

\setcounter{equation}{0}

\section{The Rotating Mass Matrix}

The Higgs boson loop $\Sigma^{\phi 1}$ in (\ref{Sigmas}) of Figure 
\ref{oneloop}(b) not only rotates the fermion mass matrix but rotates it 
in a manner which depends on the renormalization scale.  To see this, let 
us write the logarithm in (\ref{FK}) and (\ref{GK}) at any scale $\tilde{\mu}$
as a sum of its value at some given scale $\mu$ plus a scale-dependent term, 
thus:
\begin{equation}
\ln \left[\frac{m^2 x^2}{\tilde{\mu}^2}+\frac{M_K^2}{\tilde{\mu}^2}(1-x) \right]
   = \ln \left[ \frac{m^2 x^2}{\mu^2} + \frac{M_K^2}{\mu^2}(1-x) \right]
   + \ln [\mu^2/\tilde{\mu}^2].
\label{lnscale}
\end{equation}
If we change the scale from $\mu$ to $\tilde{\mu}$, then $\Sigma^{\phi 1}$ in 
(\ref{Sigmas}) will change by the amount:
\begin{equation}
\left\{ -\frac{1}{32 \pi^2} \sum_K m \gamma_4 \bar{\Gamma}_K \gamma_4 
  \bar{\Gamma}_K + \frac{1}{16 \pi^2} \sum_K \bar{\Gamma}_K m 
   \bar{\Gamma}_K \right\} \ln[\tilde{\mu}^2/\mu^2].
\label{runrotate}
\end{equation}
Recalling (\ref{Gammabar}), we can rewrite the first term within the curly
brackets as:
\begin{equation}
- \frac{1}{32 \pi^2} \sum _K m \{ \bar{\gamma}_K^\dagger \bar{\gamma}_K
   \frac{1}{2}(1+\gamma_5) + \bar{\gamma}_K \bar{\gamma}_K^\dagger
   \frac{1}{2}(1-\gamma_5) \},
\label{runrota1}
\end{equation}
and the second term as:
\begin{equation}
\frac{1}{16 \pi^2} \sum_K \{\bar{\gamma}_K m \bar{\gamma}_K 
   \frac{1}{2} (1+\gamma_5) + \bar{\gamma}_K^\dagger m \bar{\gamma}_K^\dagger
   \frac{1}{2}(1-\gamma_5) \}.
\label{runrota2}
\end{equation}
On substituting $\bar{\gamma}_K$ from (\ref{gammabar}) and summing over 
$K$ we obtain for the first term:
\begin{equation}
- \frac{1}{32 \pi^2} m_T \rho^2 \left \{3 \left( \begin{array}{c} x \\ y \\ z
   \end{array} \right) (x, y, z) \frac{1}{2} (1 + \gamma_5)
   + \left( \begin{array}{c} x \\ y \\ z \end{array} \right)
   (\tilde{x}_1, \tilde{y}_1, \tilde{z}_1) \frac{1}{2} (1 - \gamma_5) \right\},
\label{runrota1a}
\end{equation}
and for the second term:
\begin{equation}
\frac{1}{16 \pi^2} m_T \rho^2 \left\{ - \left( \begin{array}{c}
   \tilde{x}_1 \\ \tilde{y}_1 \\ \tilde{z}_1 \end{array} \right) (x, y, z) 
   \frac{1}{2} (1 + \gamma_5) - \left( \begin{array}{c} x \\ y \\ z 
   \end{array} \right) (\tilde{x}_1, \tilde{y}_1, \tilde{z}_1) 
   \frac{1}{2} (1 - \gamma_5) \right\},
\label{runrota2a}
\end{equation}
with:
\begin{equation}
\tilde{x}_1 = \frac{x(x^2-y^2)}{x^2+y^2} + \frac{x(x^2-z^2)}{x^2+z^2},
   {\rm cyclic},
\label{xtilde1}
\end{equation}
where we have kept only the contributions from $K = 3, 5, 7$ since those
from $K = 1$ and the sum of $ K = 2, 4, 6$ affect only the normalization 
of $m$, while those from $K = 8, 9$ both vanish.  In principle, of course,
a change in the normalization of $m$ will get reflected also in its
orientation, but this is of second order in smallness if the change in
scale is small and can therefore be neglected.

The scale-dependent corrections (\ref{runrota1a}) and (\ref{runrota2a})
leave the mass matrix factorized, as expected from the arguments given in 
ref \cite{Chantsou}, but no longer hermitian.  However, following the 
convention introduced above in (\ref{fermassmatp}) and (\ref{fermassmatpW}), 
one can redefine the right-handed fermion fields again so as to make the 
corrected mass matrix $m'$ hermitian.  The net result then is that, apart 
from changes in the normalization which may be ignored, one obtains from
these terms a rotation to the mass matrix which may be represented as a 
rotation to the vector $(x, y, z)$, thus:
\begin{equation}
\left( \begin{array}{c} x \\ y \\ z \end{array} \right) \longrightarrow
   \left( \begin{array}{c} \tilde{x} \\ \tilde{y} \\ \tilde{z} \end{array} 
   \right) = \left( \begin{array}{c} x \\ y \\ z \end{array} \right)
   + \frac{5}{64 \pi^2} m_T \rho^2 \left( \begin{array}{c} \tilde{x}_1 \\
   \tilde{y}_1 \\ \tilde{z}_1 \end{array} \right) \ln [\tilde{\mu}^2/\mu^2],
\label{x'y'z'}
\end{equation}
which depends on the change in scale.

\begin{figure}[htb]
\vspace{0cm}
\centerline{\psfig{figure=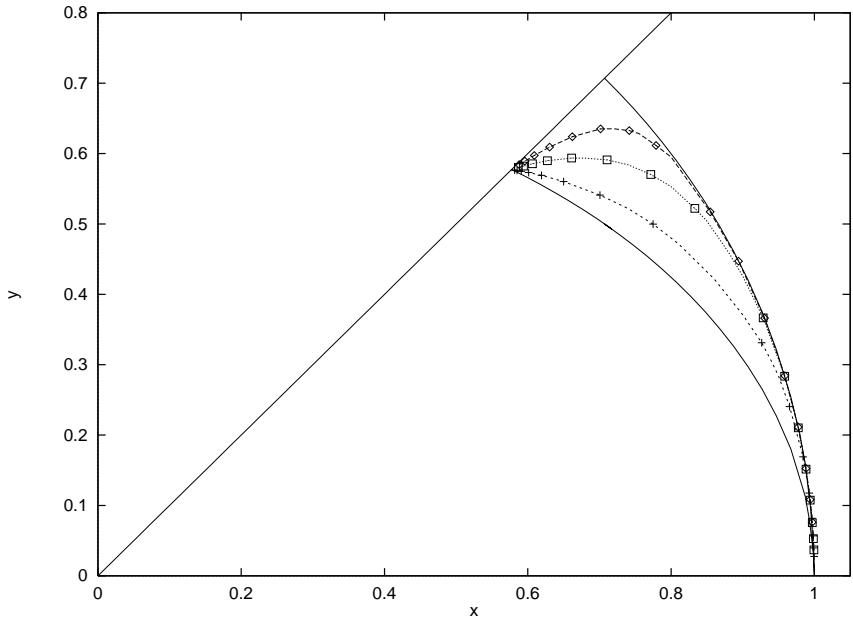,width=0.6\textwidth}}
\vspace{0cm}
\caption{The `running' of the vector $(x', y', z')$}
\label{runtraj}
\end{figure}

By iterating the formula (\ref{x'y'z'}) in small steps, one can evaluate
the rotation in $(x', y', z')$ over finite changes of scales.  One arrives
then at a picture similar to the familiar one of running coupling constants,
except that here it is a normalized vector $(x', y', z')$ that `runs'.  From
(\ref{xtilde1}), it is readily seen that for $(x', y', z')$ equal to $(1,0,0)$
or $\frac{1}{\sqrt{3}}(1,1,1)$, the increment due to a change of scale
vanishes.  These 2 vectors are thus fixed points in the usual sense under
changes of scales.  It can also be seen from (\ref{xtilde1}) that for other 
values of $(x', y', z')$ (where we have adopted the convention $x' > y' >z'$),
and for decreasing energy scales, the increment satisfies the following 
inequalities:
\begin{equation}
\frac{\tilde{x'}}{\tilde{y'}} < \frac{x'}{y'}; \ \ \ 
   \frac{\tilde{y'}}{\tilde{z'}} < \frac{y'}{z'},
\label{runineq}
\end{equation}
which mean that for decreasing scales, the vector $(x', y', z')$ will `run'
away from the fixed point $(1,0,0)$ towards, in general, the fixed point
$\frac{1}{\sqrt{3}}(1,1,1)$, tracing out a tracjectory as the scale decreases.
These assertions are confirmed by the numerical calculation presented in 
Figure \ref{runtraj}, where the spacing between points on a trajectory denotes
the speed (in arbitray units) at which the vector $(x', y', z')$ runs as a 
function of $\ln \mu$.  Since $(x', y', z')$ is normalized by definition, only
the values of $x'$ and $y'$ need be presented and the trajectories are 
bounded by the circle $x'^2 + y'^2 = 1$, while the convention adopted above,
of $x' > y' > z'$, restricts the trajectories to within the line $x' = y'$ and 
the ellipse $y'^2 = (1-x'^2-y'^2)$.  This figure gives us a very useful picture 
to which we shall often refer.

We recall that the vector $(x', y', z')$ represents actually the two 
identical factors of the factorized mass matrix (\ref{fermassmatpW}) at 
the 1-loop level, so that a rotating $(x', y', z')$ implies a mass matrix 
with a scale-dependent orientation.  Now, for such a case, it is not so 
obvious how the mass and state vector of each individual state ought to be 
defined.  This ambiguity is not a peculiarity of the dualized standard model
alone but arises already in the ordinary (i.e. nondualized) standard model
where a mass matrix with scale-dependent orientation occurs by virtue of the 
non-diagonal CKM matrix in the renormlaization group equation \cite{Chantsou};
only there, the effect being small, its consequences can be neglected.  The 
point is the following.  At any scale, the mass matrix can of course be
diagonalized and, being hermitian, its eigenvalues will be real and, if 
nondegenerate, their eigenvectors are orthogonal to one another so that the 
transformation matrix, say $S$, from the original basis to the new basis formed 
by the eigenvectors will be unitary.  However, these eigenvalues are scale 
dependent, and cannot as yet be identified as the mass of the individual 
states.  Usually, the actual mass of a state is defined as the value of 
the scale-dependent mass evaluated at the scale equal to its value at that 
scale: $m(\mu)=\mu$.  Here, however, since the eigenvalues are different by 
assumption, there is no scale for which this criterion can be satisfied 
simultaneously for all of them.  One can, of course, take each eigenvalue 
and evaluate it at the scale equal to its value, and hence define at this 
scale the corresponding eigenvector as the state vector of the particle 
with this value as its mass.  But since the orientation of the mass matrix 
is itself supposed to depend also on the scale, the state vectors so 
defined for the various particles at different scales will in general not 
be orthogonal to one another, so that the transformation $S$ from the 
original basis to this new basis of state vectors will not in general be 
unitary.  In fact, we do not know of a general prescription which can
define, from a rotating mass matrix, the masses and state vectors of the 
individual particle states by the normal criterion which yet leaves the 
transformation matrix $S$ unitary.

However, for a mass matrix which remains factorizable at all scales as
the one considered here, it turns out that there is a way \cite{Chantsou}
in which masses and state vectors can be defined in accordance to the 
normal criterion and which gives a unitary $S$ matrix.  To be specific, let 
us take first the $U$-type quarks as example.  The mass matrix being 
factorizable, there is only one non-zero eigenvalue at any scale.  Suppose 
we evaluate this nonzero eigenvalue at the scale equal to its value.  Then, 
in accordance with the standard criterion above, we can define this value 
unambiguously as the mass of the top quark.  The state vector of the top 
quark is thereby also defined uniquely as that eigenvector with the nonzero 
eigenvalue at the scale of the top mass, which in the present framework is 
just the vector $(x', y', z')$ taken at the top mass.  At this scale, the 
other two eigenvalues are zero, but they should not be regarded as 
the masses of the two lower generations for they are evaluated at the 
wrong scale.  Furthermore, one does not know at this stage which two 
vectors should correspond to the 2 lower generations.  However, being 
physically independent states, the 2 lower generations ought to have state 
vectors lying in the 2-dimensional subspace orthogonal to the top state vector, 
namely in the space spanned by two vectors with zero eigenvalues which we 
may choose as:
\begin{equation}
|{\bf v}_2 \rangle = -\beta \left( \begin{array}{c} y'-z' \\ z'-x' \\ 
   x'-y' \end{array} \right),
|{\bf v}_3 \rangle = \beta \left( \begin{array}{c} 1-x'(x'+y'+z') \\ 
   1-y'(x'+y'+z') \\ 1-z'(x'+y'+z') \end{array} \right),
\label{vsub23}
\end{equation}
with
\begin{equation}
\beta = 1/\sqrt{3-(x'+y'+z')^2},
\label{betaagain}
\end{equation}
although at this stage one does not know which linear combinations of these 
two vectors should represent the $c$ quark and which the $u$ quark.  

Suppose we now lower the scale $\mu$ by some small amount $\Delta \mu$.
Then we know from (\ref{x'y'z'}) that the mass matrix will be rotated via 
a rotation of the vector $(x', y', z')$ by a small amount proportional to 
$\ln (\mu^2/m_t^2)$.  Hence, by repeated application of this procedure, 
one can evaluate the loop-corrected mass matrix $\tilde{m}'$ at a scale 
some finite amount lower than the top mass.  At the lower scale, because 
of the rotation, the vectors $|{\bf v}_i \rangle$ are no longer eigenvectors 
of $\tilde{m}'$, so that in particular the mass submatrix:
\begin{equation}
\langle {\bf v}_i| \tilde{m}' |{\bf v}_j \rangle, \ \ \  i, j = 2, 3,
\label{msubmat}
\end{equation}
which was zero at the top mass scale is here no longer the null matrix.
However, being a submatrix of a rank 1 matrix, it is still of at most
rank 1 and hence has at most one nonzero eigenvalue, the value of which
will depend on the scale where the expression is evaluated.  Applying
then the same reasoning as for the top quark, we now vary the scale 
until the nonzero eigenvalue of (\ref{msubmat}) comes out equal to the 
scale itself, which value we shall call the charm mass for consistency.  
At the same time, the eigenvector $|{\bf v}_2' \rangle$ corresponding to 
this eigenvalue at this scale we define as the state vector of the $c$
quark, which is, of course, by definition orthogonal to the state vector
$|{\bf v}_1 \rangle$ defined above for the $t$ quark, as it should be.
Furthermore, the state vector of the $u$ quark is now also automatically 
given as the vector $|{\bf v}_3' \rangle$ which is orthogonal to both 
the top quark state vector $|{\bf v}_1 \rangle$ and the charm quark state 
vector $|{\bf v}_2' \rangle$ already defined.  At this stage, then, the 
state vectors of the 3 generations are all specified which we were unable 
to do before.  

Finally, to find the mass of the $u$ quark, we lower the scale again in small 
steps, applying repeatedly (\ref{x'y'z'}), to some scale.  This scale we then
vary until the value of
\begin{equation}
\langle {\bf v}_3'| \tilde{\tilde{m}}' |{\bf v}_3' \rangle 
\label{mp33}
\end{equation}
becomes equal to the scale itself, and this we define as the mass of the
$u$ quark, again in conformity to the procedure above.  So now the masses
of all three physical states are also defined, and they will all in
general be nonzero.

We notice that the masses of all three generations $t, c, u$ here are 
each defined using the normal criterion of evaluating the appropriate 
eigenvalue of the mass matrix at the scale equal to its value.  Moreover, 
the three state vectors corresponding to the three generations so defined 
are also mutually orthogonal so that the matrix $S$ transforming from the 
original ``gauge basis'' to the ``physical basis'' of state vectors is 
unitary as it ought to be.  The actual values of t he masses and state 
vectors so defined depend on the manner that the mass matrix rotates
as a function of the energy scale, which in our present scheme depends in
turn on the vacuum expectation values $x, y, z$ of the (dual colour) Higgs 
fields and on the strength $\rho_U$ of their Yukawa couplings to the 
$U$-type quarks, the values of which parameters have yet to be specified.

A similar procedure applied to the $D$-type quarks defines in turn the 
masses and state vectors of the $b, s$, and $d$ quarks.  The actual values 
of these quantities in the present scheme will depend on the same Higgs 
fields vacuum expectation values $x, y, z$ as for the $U$-type quarks but 
in general a different Yukawa coupling strength $\rho_D$.  Together with 
$m_T$, the normalization of the mass matrix for each quark type $T$, which 
may be identified with the highest generation mass, there are altogether 
then 6 parameters, namely $m_U$, $m_D$, $\rho_U$, $\rho_D$, and the vector
$(x, y, z)$ which, being normalized to unit length, counts only as 2 
parameters.  With the remaining 4 parameters then, one is required 
to evaluate the 4 masses of the 2 lower generations $m_c, m_u, m_s, m_d$ 
by the method described above, as well as the 4 relevant parameters of 
the CKM matrix in the manner outlined below.

By definition, the CKM matrix is the matrix giving the relative orientation
of the physical state vectors of the 3 $U$-type quarks to those of the 3
$D$-type quarks.  In terms of the notation introduced above, it is given
as the matrix:
\begin{equation}
V_{ij} = \langle {\bf v'}_i|{\bf v'}_j \rangle,
\label{Vij}
\end{equation}
where $i(j) = 1, 2, 3$ denote respectively $t(b), c(s), u(d)$, which in 
usual convention are arranged in the reversed order.  Now, in the literature,
the CKM matrix is often defined also as the overlap $UD^{\dagger}$ between 
the matrix $U$ which diagonalizes the mass matrix of the $U$-type quarks 
and the matrix $D$ which diagonalizes the mass matrix of the $D$-type quarks.  
This definition is equivalent to that adopted above in terms of the physical
states when the mass matrices do not rotate with the energy scale.  When
the mass matrices have scale-dependent orientations, however, the 2 
definitions differ, since the vectors which diagonalize the mass matrices, 
as explained above, need no longer represent the physical states.  Indeed,
since the mass matrices are scale-dependent so will be their diagonalizing
matrices $U$ and $D$, and so also will be the CKM matrix which is defined 
as their overlap.  On the other hand, the physical state vectors defined 
in the preceding section for the 3 generations of both the $U$-type and 
$D$-type quarks are all scale-independent, so that the CKM matrix 
defined as the transformation between the $U$ physical basis to the $D$
physical basis is also scale-independent.  Here we shall evaluate the CKM 
matrix defined as the transformation matrix (\ref{Vij}) between bases of 
physical states, which definition accords more with the philosophy adopted 
in this paper and seems to us also to correspond more to what is actually 
measured experimentally.  

Before we proceed to numerical work, however, let us first note a qualtitaive
feature of the present procedure which is of relevance both to our future 
calculation and to its comparison with experiment.  The empirical CKM matrix,
though near identity, has off-diagonal elements differing considerably in 
size, varying from around 20 percent for $V_{cd}$ and $V_{us}$ through a
few percent for $V_{ts}$ and $V_{cb}$ to just a few permille for $V_{td}$ 
and $V_{ub}$.  This variation may seem difficult to explain since if the
matrix is due to some effect rotating the $U$-type mass matrix relative
to the $D$-type mass matrix, one would expect the mixing elements to be of 
roughly the same order of magnitude.  In the present scheme, however, there 
is a natural explanation for this variation.  We recall that the state vectors 
of the two lower generations are to be defined through the running rotation 
of the mass matrices, so that these vectors get an extra kick in orientation 
in addition to that of the frames at the top and bottom mass.  And it is 
this effect, having strictly to do with the special way that the physical 
states of the lower generations are here defined, which gives the Cabibbo 
angle a sort of special status among CKM matrix elements and hence, as we 
shall see, a particularly large value in comparison with the others as 
experimentally observed.

\setcounter{equation}{0}

\section{Numerical Results}

To perform the calculation outlined in the preceding section, given any
vector $(x, y, z)$ for the Higgs fields vacuum expectation values which 
also doubles as the factor of the zeroth order fermion mass matrix, we 
face as our first task the evaluation of the 1-loop corrected vector:
\begin{equation}
|{\bf v}_1 \rangle = \left( \begin{array}{c} x' \\ y' \\ z' \end{array} \right)
   = \left( \begin{array}{c} x \\ y \\ z \end{array} \right)
   - \left( \begin{array}{c} x_1 \\ y_1 \\ z_1 \end{array} \right),
\label{v1again}
\end{equation}
(properly normalized), for $x_1, y_1, z_1$ evaluated at the energy scale,
respectively for the $U$- and $D$-type quark, of the top and bottom mass.  
This cannot be done by applying directly the formula (\ref{x1phi1}) derived 
above for the following reason.  The expression (\ref{x1phi1}) depends on 
the masses of the Higgs bosons $M_K$, which in turn depend on the strength 
$\zeta$ of the Higgs vacuum expectation values and the Higgs self-couplings 
$\kappa$ and $\lambda$.  Of these parameters, $\lambda$ is irrelevant since 
it occurs only in $M_1$ entering in $A_1$ and $B_1$ of (\ref{x1phi1}) which 
are seen to affect only the normalization of $(x_1, y_1, z_1)$, not its 
orientation, while the other two occur only together in the combination 
$\kappa \zeta^2$, as seen in (\ref{MK1to7}).  From the lower bound on the 
dual gauge boson mass of around 400 TeV deduced from the absence of 
flavour-changing neutral currents effects in meson decays, one obtains 
from (\ref{muNdiag}) an estimate of about 20 TeV for a lower bound on 
$\zeta$.  Assuming the coupling $\kappa$ to be of order unity, this gives 
an estimate for a lower bound on $M_K$ also of around 20 TeV.  Now the 
formula for $(x_1, y_1, z_1)$ involves $\ln M_K^2$, which when evaluated 
directly at the top and bottom mass scales as we desired would be very 
large and violate the spirit of our present perturbative calculation.
However, there is no real problem in this, for we can always evaluate first
the correction $(x_1, y_1, z_1)$ at the scale of the Higgs mass, say 20 TeV,
and then use the formula (\ref{x'y'z'}) to `run' the corrected vector by 
steps down to the scales of the top and bottom mass.  At every step, then,
the calculation would be perturbative for the correction is kept always 
small.  This is in the spirit of the original Gell-Mann-Low idea 
\cite{Gellmanlow} which led to the renormalization group equation. 

A calculation done in this way, however, still leaves it dependent explicitly
on the masses of the Higgs bosons.  This would be a little awkward but 
for a happy and quite intriguing `accident' to be elucidated later, for 
these masses are known only by the tree-level formulae (\ref{MK1to7}) which 
are likely to be strongly renormalized, like the fermion masses, by e.g. the 
dual gluon loops.  Because of this `accident', however, it turns out that
to a very good approximation we can put all the Higgs boson masses equal,
say to a common scale $M_H$, even the value of which in the end does not 
really matter, but which we take at the moment to be 20 TeV.  We need then 
to evaluate the formulae (\ref{x1phi1}) for the common scale $M_H = M_K$ 
for all $K$.  This expression is almost the same for $U$- and $D$-type 
quarks, and indeed even for leptons, differing only in the normalization 
$m_T$ of the mass matrix.  This difference is small, only of the order of 
$m_T^2/M_H^2$ which for $M_H$ around 20 TeV, is less than $10^{-4}$, as we 
have checked both by analytic and numerical calculations.  It can thus 
be safely neglected.  This means that whatever the correction due to 
(\ref{x1phi1}) happens to be at the scale $M_H$ (which is in fact quite small 
numerically), it will be the same for all the fermions.  Therefore, in 
the present approximation when all $M_K$'s are put equal, we can just start 
at the scale $M_H$ with the same values of $x', y', z'$ for all fermions, 
and simply `run' them down to the mass of the highest generation for each 
fermion type to evaluate the vector $|{\bf v}_1 \rangle$ in (\ref{v1again}) 
for each case.

The `running' mechanism (\ref{x1phi1}) and the starting values at $M_H$ both
being the same for all fermion types, the vector $(x', y', z')$ will in this
approximation `run' along the same trajectory, only possibly at different
speeds because the Yukawa coupliongs $\rho_T$ may be different.  In any case, 
since the $m_T$'s are different, one would arrive at a different state vector
for the highest generation $|{\bf v}_1\rangle$ for the different fermion 
types.  Starting with some input values for $\rho_T$ and some initial 
values for $(x', y', z')$, say $(x_I, y_I, z_I)$, at the scale $M_H$, 
and applying (\ref{x1phi1}) repeatedly in small steps, one would arrive
at some value for $|{\bf v}_1 \rangle$ for each fermion type.  In principle,
the steps should be infinitesimal, but in our numerical calculation we used
typically about 500 steps for each decade of energy which we found were
just about sufficient for the 1\% accuracy that we wanted.

Having defined $|{\bf v}_1 \rangle$ for each fermion type, we can now `run'
further down in energy scale to the second generation mass.  As the mass
matrix rotates in running, the mass will `leak' into the second generation
and give it a mass, as explained in the section above.  The amount of 
leakage will depend on the value of the Yukawa coupling strength $\rho_T$
and the range of energy run.  Hence, the mass obtained from leakage at
the mass scale of the second generation will in general be different for
the different fermion types and different also from the actual input mass
of the second generation.  By adjusting the values of $\rho_T$, one can
adjust the amount of leakage and hence ensure that the leaked mass
obtained for the second generation is indeed the same as the input mass
for each fermion type.  Let us call these optimized values of $\rho_T$ so 
obtained at this stage as the output $\rho$'s.  

These output $\rho's$, however, were determined starting from some vector
$|{\bf v}_1 \rangle$ for the highest generation, which in turn depended
on the assumed input values of $\rho_T$ used to run the initial vector
$(x_I, y_I, z_I)$ from the scale $M_H$ down to the scale of the highest
generation.  Obviously, the input and output values of these $\rho$'s
need not be the same.  We have thus to optimize again and adjust the
input values of $\rho_T$ until the output value is in each case the same
as the output value of $\rho_T$ obtained from it.  This optimzed value
we now call the fitted $\rho_T$.

With these fitted values for $\rho_T$ giving good second generation masses, 
we can now determine the state vectors $|{\bf v'}_i \rangle, i=2,3$ both 
for the second and the lowest generations, as we explained in the preceding
section.  Then, with the physical state vectors for all three generations 
and both $U$- and $D$-type quarks all determined, the CKM matrix easily 
follows from (\ref{Vij}).  Further, by running down to even lower energy 
scales, we can calculate the mass of the lowest generation by requiring 
that the `leaked' mass in the lowest state in some scale be equal to the 
scale itself to which it is run.  One sees thus that given any initial
value $(x_I, y_I, z_I)$ at the scale $M_H$, our program automatically 
determines for us the values of $\rho_T$ which fit the masses of the top
2 generations for each fermion type, and then gives us the CKM matrix and
the lowest generation masses as the result.  We have thus in effect just 
2 real parameters left to adjust with which to calculate all these physical 
quantities.  

We recall that the present scheme does not allow us to calculate the
absolute normalization of the mass matrix, nor its variation with the 
energy scale, but only the orientation of the mass matrix.  We are 
therefore more confident with our result on the CKM matrix which depends
only on the orientation than on the fermion masses.  The calculation of
the fermion masses depends in principle on the scale-dependence not only
of the normalization of the mass matrix but also of the $\rho$'s, which
dependence, for lack of anything better, we have simply ignored.  In our 
calculation therefore, we have concentrated on getting a good fit to the 
CKM matrix rather than the masses of the lowest generation.

With the whole calculation involving only real quantities, it is clear
that we shall not be able to obtain any CP-violating phase in our CKM
matrix.  There are thus only 3 independent real parameters in the CKM 
matrix to calculate.  We focus first our attention on the last row and
column of $V$, namely that labelled by $t$ and $b$.  The state vectors
of $t$ and $b$, which in our notation were denoted by $|{\bf v}_1^U \rangle$
$|{\bf v}_1^D \rangle$ respectively, are not affected by the additional 
rotation of the physical states from the highest to the second generation, 
which, as explained in the last paragraph of the preceding section, is 
responsible for the particularly large value of the Cabibbo angle.  Their
relative orientation therefore give the measure of the relative rotation
of the vectors $(x', y', z')$ when run from the starting value at $M_H$ to
the respective highest generation mass.  One sees that the difference in
orientation between these two states are quite small, the off-diagonal 
elements being only of the order of a few percent in magnitude.  However, 
the distances run from the starting point $M_H$ to respectively the $t$ 
and $b$ mass are quite different, being only about 2 decades in energy 
for the $t$ and nearly 4 decades for the $b$.  Therefore, to end up 
with only a few percent difference in orientation, either the the Yukawa 
couplings $\rho_T$ must be so small as to give little running, which would 
contradict the sizeable amount of `leakage' required to give the second 
generation mass, or the vectors $(x', y', z')$ have to be near a fixed 
point so that the running is rather inefficient.  We explored first the 
`upper' fixed point $\frac{1}{\sqrt{3}} (1,1,1)$, but found no sensible 
solution.  The `lower' fixed point $(1,0,0)$, however, proved productive.

In the input initial values of $(x_I, y_I, z_I)$, for $x_I \sim 1$ and
$y_I > z_I$ but both small, it is, crudely speaking, $y_I$ which tells us 
how far down we are on the trajectory, and the relative size of $z_I$ to
$y_I$ which tells us which trajectory we are on.  By adjusting $y_I$,
one can thus make the relative orientation between the $t$ and $b$ states,
as exhibited in e.g. $V_{ts}$ and $V_{cb}$, to be of the order of a few
percent as required by experiment.  Then, by adjusting $z_I$, to which
$V_{ts}$ and $V_{cb}$ are quite insensitive, one can fit the Cabibbo
angle $V_{us}$ and $V_{cd}$ to the empirical value of around 20 percent.
The best result we have obtained so far in this exercise is shown below:
\begin{equation}
|V_{rs}| = \left( \begin{array}{lll} 
   0.9755 & 0.2199  & 0.0044 \\ 
   0.2195 & 0.9746  & 0.0452 \\
   0.0143 & 0.0431  & 0.9990
   \end{array} \right),
\label{CKMout}
\end{equation}
This is to be compared with the result below obtained from experiment
\cite{databook}:
\begin{equation}
|V_{rs}| = \left( \begin{array}{lll}
   0.9745 - 0.9757 & 0.219 - 0.224  & 0.002 - 0.005 \\
   0.218 - 0.224  & 0.9736 - 0.9750 & 0.036 - 0.046 \\
   0.004 - 0.014  & 0.034  - 0.046  & 0.9989 - 0.9993
   \end{array} \right).
\label{CKMexpt}
\end{equation}
The agreement is seen to be good.  This we find encouraging, first that
we can indeed adjust our parameters to obtain good values for the Cabibbo
angle and the $V_{ts}$ and $V_{cb}$ elements, and second that once we 
have fitted these to approximately the right values, then $V_{ub}$ and 
$V_{td}$ automatically come out to be a few permille in magnitude as 
experimentally observed, which seem to indicate that the method we used 
for defining the lower generations states have somehow got the orientation 
right.  We have calculated also with the same values of the parameters 
certain products and ratios of matrix elements which have been independently 
measured, obtaining: 
\begin{eqnarray}
|V_{ub}|/|V_{cb}| & = & 0.0983, \nonumber \\
|V_{td}|/|V_{ts}| & = & 0.3310, \nonumber \\
|V_{tb}^{*} V_{td}| & = & 0.0142,
\label{Vratioout}
\end{eqnarray}
to be compared with the values below quoted from the databook \cite{databook}:
\begin{eqnarray}
|V_{ub}|/|V_{cb}| & = & 0.08 \pm 0.02, \nonumber \\
|V_{tb}|/|V_{ts}| & < & 0.37, \nonumber \\
|V_{tb}^{*} V_{td}| & = & 0.009 \pm 0.003,
\label{Vratioexp}
\end{eqnarray}
The agreement is again reasonable.

These numbers were calculated with the following masses in GeV for the 
fermions in the 2 highest generations:
\begin{equation}
m_t = 176, \ m_b = 4.295, \ m_\tau = 1.777, \ m_c = 1.327, \ m_s = 0.173,
   \ m_\mu = 0.106,
\label{mfortetc}
\end{equation}
where the central value was taken where such is given but otherwise the 
geometric mean of the upper and lower experimental bounds as listed in 
the databook \cite{databook}.  We have included in the fit the charged 
leptons $\tau$ and $\mu$ which, though not entering into the CKM matrix, 
can be dealt with in the same manner as the quarks at the cost of only
an extra $\rho$ parameter.  The initial values of $(x_I, y_I, z_I)$
at the scale of $M_H$ = 20 TeV chosen to fit the CKM elements were:
\begin{equation}
x_I = 0.999998, \ y_I = 0.002200, \ z_I = 0.000025.
\label{xIyIzI}
\end{equation}
The fitted $\rho$'s which emerged automatically from the requirement of
consistency with the input masses (\ref{mfortetc}) turned out then to be:
\begin{equation}
\rho_U = 3.4737, \ \rho_D = 3.3693, \ \rho_L = 3.4728,
\label{rhovalues}
\end{equation}
which are encouragingly all of order unity.

One quite amazing feature of the parameters obtained from the fit is the close 
proxity to one another of the values of the Yukawa couplings $\rho$ for all 
three fermion types, the spread of which in (\ref{rhovalues}) is only around 
1.5 parts per mille.  The actual values listed in (\ref{rhovalues})
depend of course on the input values of the masses (\ref{mfortetc}) of 
the fermions of the 2 higher generations.  However, even if we vary these 
input masses to the utmost extremes allowed by the experimental bounds, 
the $\rho$'s are found by calculation to remain roughly equal, differing 
from one another always by less than 10 percent.  At first sight, this 
may seem strange for the ratio of the second to highest generation mass
differ considerably from fermion type to fermion type.  For example,
$(m_c/m_t) \sim$ 0.7 percent, while $(m_s/m_b) \sim$ 4 percent, a factor 
of 6 different, which would mean that the `leakage' of mass by running 
from $b$ to $s$ must be several times stronger than that from $t$ to $c$,
suggesting that the coupling $\rho$ which governs the speed of this
running ought to be several times bigger for the $D$-type than for the 
$U$-type.  The reason why this does not happen in the present calculation
is that, the $t$ quark being heavier than the $b$, lies further down the
trajectory depicted in Figure \ref{runtraj}, i.e. nearer the fixed point
$(1,0,0)$, so that the running rotation there is much less efficient than
at the $b$ mass which lies much higher on the trajectory.  Hence, with
about the same value for $\rho$ one can still obtain widely different 
`leakages' in the two cases.  However, that the fitted values of 
$\rho_T$ should come out so close to one another for all 3 fermion
types is a bit of a surprise.

This approximate equality of the $\rho$'s for all 3 fermion types is what
we called our ``happy accident'' at the beginning of this section which
gives us a number of practical advantages in our calculation.  First, we 
recall that in the calculation reported above, we had made the simplifying 
assumption that all the Higgs bosons had the same mass $M_H$, which would 
be far from the truth if we believe the tree-level relations (\ref{MK1to7}) 
given the very different values we need in (\ref{xIyIzI}) for $x, y$ and $z$.  
To take Higgs mass splitting into account, one ought in principle to proceed
as follows.  One first goes up to the scale of the highest Higgs mass, 
which in the present case, according to (\ref{MK1to7}) and (\ref{xIyIzI}), 
would be $M_4 \sim M_6 \sim$ several orders of magnitude higher than the
lowest Higgs mass $M_H$ of around 20 TeV.  At this high scale, we have 
next to calculate the rotation to the original Higgs vacuum expectation 
values $(x, y, z)$ due to the $K = 4,5,6,7$ terms in the Higgs loop 
diagram (\ref{x1phi1}), and then run the resulting $(x', y', z')$ down 
to the scale of the lightest Higgs, namely $M_2 = M_H$.  Then the result 
of this running has to be added to result of the rotation to the
original $(x, y, z)$ due to the $K=2,3$ terms in (\ref{x1phi1}), and it 
is this sum that we have in principle to use as the intitial vector 
$(x_I, y_I, z_I)$ for our above calculation.  If the Yukawa couplings 
$\rho$ were different for the 3 fermion types, then $(x_I, y_I, z_I)$ so 
obtained would be different also.  Now, however, because of the ``happy 
accident'' of the $\rho$'s being the same, (and the $m_T$-dependence of 
(\ref{x1phi1}) being, as explained before, negligible), the resulting 
$(x_I, y_I, z_I)$ of the above manoeuvre would be the same for the 3 
fermion types.  Hence, our `simplifying' assumption made at the beginning 
of starting with the same $(x_I, y_I, z_I)$ for all fermion types at scale 
$M_H$ is now {\it a posteori} entirely justified.

Further, this ``happy accident'' implies also that the calculation is
actually independent of the scale $M_H$ which we have so far chosen to be
20 TeV.  To see this, recall that we were supposed to start with some 
$(x_I, y_I, z_I)$ at $M_H$ for all fermion types and run the vector, with 
the appropriate $\rho$'s, down to respectively the $t$, $b$ and $\tau$ 
mass values.  If the $\rho$'s were different, then starting with a
different $M_H$, one would arrive at different values at the highest
generation mass for the 3 fermion types.  Now that the $\rho$'s are the
same, however, there is only one value of $(x', y', z')$ at every point of
the trajectory.  One can therefore start at any point of the trajectory
with some $(x_I, y_I, z_I)$ for all 3 types of fermions and obtain the
same answer.  That this assertion holds even for approximately equal
$\rho$'s has been checked numerically by repeating our calculation for
various starting points $M_H$.  It means that $M_H$ can be removed
altogether from our calculation as a parameter, leaving thus only the
2 ratios between $(x_I, y_I, z_I)$ as the only parameters in the calculation, 
as we have claimed. 

The other intriguing feature of the fit is the proximity of the fitted 
values in (\ref{xIyIzI}) of these $(x_I, y_I, z_I)$ to the fixed point 
$(1,0,0)$.  In contrast to the approximate equality of the $\rho$'s 
discussed above, this outcome is no accident but, as already explained before, 
is required by the smallness of all other off-diagonal CKM matrix elements 
besides the Cabibbo angle.  Although the values of $(x_I, y_I, z_I)$ at the
arbitrary starting point $M_H$ do not by themselves have much significance,
we can deduce from them the vacuum expectation values $(x, y, z)$ of the 
Higgs fields by running the scale backwards up to the highest Higgs mass 
and evaluating (\ref{x1phi1}) there.  Assuming the lowest Higgs mass 20 TeV,
the tree-level formulae (\ref{MK1to7}), and using (\ref{x1phi1}), one obtains 
in this way for the vacuum expectation rough values of the Higgs fields:
\begin{equation}
x \sim 1, \ \ y \sim 5 \times 10^{-5}, \ \ z \sim 1 \times 10^{-8},
\label{higgsvevs}
\end{equation}
which are very close indeed to the fixed point $(1,0,0)$

Though perhaps just fortuitous, the proximity of the fitted $(x, y, z)$ 
to the fixed point $(1,0,0)$ and the near equality of the fitted $\rho$'s 
are so remarkable that it is tempting to consider the exciting possibility 
of the coincidence representing in fact a symmetry which is exact in some 
approximation and is only perturbed from it by an external agent.  One 
possibility, for example, could be that if electroweak effects are neglected, 
then $(x, y, z)$ is exactly $(1,0,0)$ and the $\rho$'s are exactly equal, 
and it is only the electroweak effects which give rise to the quantities's 
departure from the equilibrium values.  At this stage, of course, the 
suggestion is a pure conjecture on our part, but it may be a worthwhile 
conjecture to entertain.

\begin{figure}
\vspace{-2cm}
\centerline{\psfig{figure=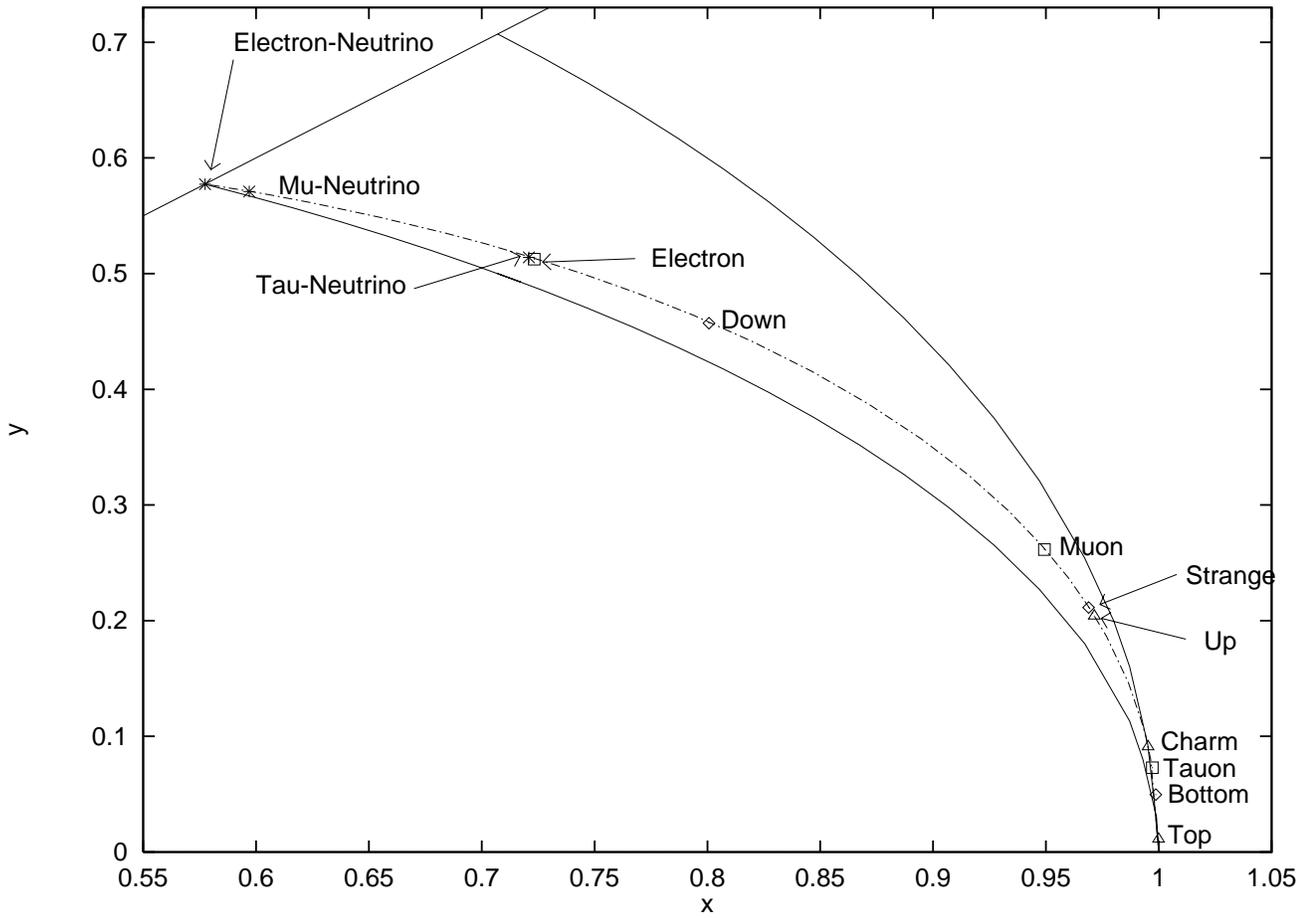,width=0.9\textwidth}}
\vspace{1cm}
\caption{The locations of the various fermion states on the common trajectory}
\label{trajloc}
\end{figure}

Having now determine the parameters of the problem, it is an easy matter
to run the vector $(x', y', z')$ further down in the energy scale and
evaluate the masses of the lowest generation fermions following the method
outlined in the preceding section.  Notice, however, that this calculation
depends in principle on the scale-dependence both of the normalization of
the mass matrix and also of the $\rho$'s, neither of which are calculable 
in the present framework.  In fact, even the earlier calculation of the
CKM matrix depends to some extent on these through fitting the $\rho$'s
to the 2 higher generation fermion masses, but there, the change in scale
not being too large, the change in normalization can be masked by adjusting
the parameters in the fit, and hence not too noticeable.  In calculating
further the lowest generation masses, the effect of the neglect is
compounded, and not too good results can be expected.  Our calculation,
using the values of the fitted parameters (\ref{xIyIzI}) and (\ref{rhovalues})
and the same higher generation masses (\ref{mfortetc}) and assuming constant
$\rho$'s and mass matrix normalizations throughout the whole energy range 
of over 6 decades, gives:
\begin{equation}
m_u = 209 \ {\rm MeV}, \ m_d = 15 \ {\rm MeV}, \ m_e = 6 \ {\rm MeV}.
\label{mforude}
\end{equation}
These mass values are fairly stable with respect to variations of the input
masses for the 2 higher generations.  For variations between the ranges
given in the databook \cite{databook}, the values obtained for the lowest
generation lie in the range:
\begin{equation}
m_u = 120-360 \ {\rm MeV}, \ m_d = 12-22 \ {\rm MeV}, \ m_e = 5-11 \ {\rm MeV}.
\label{m3range}
\end{equation}
Apart from the mass of the $u$-quark, we regard these result as sensible
given the crudeness of the assumption of no scale-dependence at all for
either the Yukawa couplings $\rho$ or the normalization of the mass matrix.  
It is perhaps interesting to understand technically why the mass for the 
$u$-quark turns out to be so much worse than in the other two cases.  As 
explained above, the approximate equality of the $\rho$'s means that all 
3 fermion types lie on the same trajectory, only differing in where the 
various physical states are placed.  For the calculation here, these 
positions are shown in Figure \ref{trajloc}.
We notice there that the $t$-quark, being the heaviest fermion, lies of
necesssity the lowest on the trajectory, while the $b$ and $\tau$ both
lie higher up.  For this reason, as already explained above, the running
efficiency is much lower around the $t$ mass than for the others so that
even with the same value of $\rho$, the leakage from $t$ to $c$ is much
smaller than that from $b$ to $s$ or from $\tau$ to $\mu$.  For the run
from the second generation to the lowest, however, the $U$-type quark is 
now in the part of the trajectory where the running efficiency is high, 
while for the $D$-type quark and charged lepton, the lowest generation
is already pressing a little against the upper fixed point $\frac{1}{\sqrt{3}}
(1,1,1)$ and losing running efficiency.  Hence, we have the unwelcome large
mass for the $u$-quark but not so large for $d$ and $e$.  When the 
scale-dependence of $\rho$ and the normalization of the mass matrix are
properly accounted for, a possible scenario may be that $(x',y',z')$ will
run faster along the trajectory so that all the lowest generation states
will press against the upper fixed point $\frac{1}{\sqrt{3}} (1,1,1)$
and give lower masses for all of them, in particular for the $u$-quark.  
The investigation of this possibility, however, is beyond the scope of
the present paper.

Finally, just as a matter of curiosity, let us apply the same sort of 
reasoning to the neutrino masses also.  Assuming the same value of $\rho$
for neutrinos as for the charged leptons, we can then in principle calculate
the masses of all the neutrinos given any one of them.  Or else, given the
experimental upper bound of any neutrino, we can obtain upper bounds on
the others.  The strongest bounds obtained in this way, we found, comes
from inputting the experimental bound $<$ 0.17 MeV for the $\nu_\mu$ mass
quoted in \cite{databook}.  Using the same fitted values of $(x_I,y_I,z_I)$ 
in (\ref{xIyIzI}) we obtained:
\begin{equation}
m_{\nu_e} < 5 \ {\rm eV}, \ \ m_{\nu_\tau} < 6 \ {\rm MeV},
\label{numasses}
\end{equation}
both of which, interestingly, are stronger than the experimental bounds:
$m_{\nu_e} < 10-15$ eV, $m_{\nu_\tau} < 24$ MeV given in \cite{databook}.  
We note that in Figure \ref{trajloc}, the points representing the neutrinos 
all press quite tightly against the upper fixed point, especially for 
$\nu_e$, which is why it gets such a stringent limit on its mass.  These 
limits, however, should not be taken too seriously, since for the neutrinos, 
and indeed even for the charged leptons, there is much more than just the 
masses to be understood.

\setcounter{equation}{0}

\section{Conclusion and Remarks}

In this paper, we set out to address the question whether the Dualized 
Standard Model scheme suggested in \cite{Chantsou} is capable of giving 
reasonable CKM matrix elements and quark masses.  The question has now, we
think, been answered in the affirmative.  Not only has one been able to 
fit the masses of the 2 higher generations sensibly with Yukawa coupling
strengths all of order unity, but also with only 2 parameters then left
to fit the absolute values of CKM matrix elements very well and give 
sensible estimates as well for the fermion masses of the lowest generation 
except for the $u$-quark.  This may not mean, of course, that the scheme 
is correct, but it is at least encouraging.

The calculation was done with dual Higgs and gauge boson masses consistent 
with existing bounds obtained from flavour-changing neutral currents decays
\cite{Bordesetal,Cahnrari}, namely $\sqrt{\kappa} \zeta =$ 20 TeV, meaning
Higgs masses of order several 10's of TeV and higher and gauge boson masses of 
order several 100's of TeV and higher.  An estimate of these masses from the 
calculation, if available, would be of phenomenological significance, since 
it enters in FCNC decays \cite{Bordesetal,Bordesetalb}, and possibly also 
in understanding air showers from cosmic rays with energy greater than 
$10^{20}$ eV, namely those beyond the GZK spectral cutoff \cite{Greisemin,Boratav,Bordesetal,Bordesetala}.  
However, unfortunately for this purpose 
though fortunately for the calculation, it turns out that the calculation 
is almost independent of the dual colour Higgs and gauge boson masses 
provided that they are large, so that no useful estimate for them can yet 
be made.

The calculation gave also a rather intriguing picture of how CKM mixing and
lower generation fermion masses are generated, namely in terms of `running'
trajectories and fixed points.  Two unexpected bonuses are the close
proximity of the Higgs vacuum expectation values $(x, y, z)$ to the fixed
point $(1,0,0)$ and the near equality of the Yakawa coupling strengths
$\rho$ for the different fermion types.  If one could find a theoretical
reason why the $\rho$'s should be equal, or how $(x, y, z)$ is given that
miniscule departure from the fixed point $(1,0,0)$, one would be approaching
a fit with a single parameter (the common $\rho$) to all CKM mixings and 
fermion mass splittings, which would be fantastic.

Of outstanding problems, we have identified two.  One concerns the CP-violating
phase in the CKM matrix which, as explained already, cannot be obtained in
the present approach, at least not in first order.  The other concerns 
the special properties of the mass matrix with scale-dependent orientation, 
also already discussed in section 5.  The problem is that there does not 
seem to be a basis of state vectors with well-defined masses for which 
the mass matrix is exactly diagonal.  In fact, this problem already figures 
in the ordinary (non-dualized) formulation of the Standard Model where a 
scale-dependent orientation is induced by the CKM matrix which cannot be 
made diagonal simultaneously with the mass matrix.  The only difference is
that the effect there is quite small, as shown in \cite{Chantsou}, and is 
for that reason often ignored.  It seems to us that whichever description 
one chooses to adopt, whether in terms of the diagonal basis or the basis 
with definite masses, the physics ought to be equivalent.  However, the 
relation between the two descriptions and the physical consequences this 
implies have not been properly worked out.

\vspace{1cm}

\noindent {\large {\bf Acknowledgement}}

\vspace{.3cm}

One of us (JB) acknowledges support from the Spanish Government on
contract no. CICYT AEN 97-1718, while another (JP) is grateful to
the Studienstiftung d.d. Volkes and the Burton Senior Scholarship of
Oriel College, Oxford for financial support.  In addition, we all thank 
George Kalmus for some timely practical help which made this collaboration
possible.

\end{document}